\documentclass[latex]{cas-dc-template}

\usepackage{float}
\usepackage[numbers]{natbib}

\usepackage{subfig}
\usepackage{mwe}
\usepackage{graphicx}
\usepackage{mathtools}
\usepackage{amssymb}

\usepackage{verbatim}
\usepackage{xcolor}
\usepackage{ulem} 

\usepackage{float}
 
\usepackage{booktabs}

\usepackage[utf8]{inputenc}
\usepackage{booktabs}

\usepackage{float}

\def\tsc#1{\csdef{#1}{\textsc{\lowercase{#1}}\xspace}}
\tsc{WGM}
\tsc{QE}
\tsc{EP}
\tsc{PMS}
\tsc{BEC}
\tsc{DE}
\usepackage{float}
\begin{document}
\let\WriteBookmarks\relax
\def\floatpagepagefraction{1}
\def\textpagefraction{.001}
\shorttitle{Realtime Mobile Bandwidth and Handoff Predictions in 4G/5G Networks}
\shortauthors{Lifan Mei et~al.}

\title [mode = title]{Realtime Mobile Bandwidth and Handoff Predictions in 4G/5G Networks}                      



                        
\author[1]{Lifan Mei}[auid=000, bioid=1, orcid=0000-0003-1188-481X]

\cormark[1]
\ead{lifan@nyu.edu}
\ead[url]{lifan@nyu.edu}


\address[1]{NYU WIRELESS, Department of Electrical and Computer Engineering, New York University, Brooklyn, NY 11201, USA}

\author[1]{Jinrui Gou}

%


\address[2]{Department of Computer Science, New York Institute of Technology, New York, NY 10023, USA}

\author%
[1]
{Yujin Cai}

\author%
[2]
{Houwei Cao}

\author%
[1]
{Yong Liu}
\ead{yongliu@nyu.edu}






\begin{abstract}
Mobile apps are increasingly relying on high-throughput and low-latency content delivery, while the available bandwidth on wireless access links is inherently time-varying. The handoffs between base stations and access modes due to user mobility present additional challenges to deliver a high level of user Quality-of-Experience (QoE). The ability to predict the available bandwidth and the upcoming handoffs will give applications valuable leeway to make proactive adjustments to avoid significant QoE degradation. In this paper, we explore the possibility and accuracy of realtime mobile bandwidth and handoff predictions in 4G/LTE and 5G networks. Towards this goal, we collect long consecutive traces with rich bandwidth, channel, and context information from public transportation systems.  We develop Recurrent Neural Network models to mine the temporal patterns of bandwidth evolution in fixed-route mobility scenarios. Our models consistently outperform the conventional univariate and multivariate bandwidth prediction models. For 4G \& 5G co-existing networks, we propose a new problem of handoff prediction between 4G and 5G, which is important for low-latency applications like self-driving strategy in realistic 5G scenarios. We develop classification and regression based prediction models, which achieve more than 80\% accuracy in predicting 4G and 5G handoffs in a recent 5G dataset.
\end{abstract}




\maketitle

\section{Introduction}

The growth of mobile Internet traffic has accelerated in the recent years, thanks to both the breakthroughs in wireless access technologies, such as mmWave and massive MIMO, and a fast-growing array of mobile multimedia apps, ranging from video streaming/conferencing, Virtual Reality, Augmented/Mixed Reality, to autonomous driving, etc.  While the next-generation mobile access infrastructure, such as 5G network, is designed to deliver high-throughput, low-latency, and high-reliability, the actual Quality-of-Service delivered to users is still vulnerable to various impairments to the physical channel quality between user devices and access points. It is well-known that wireless signals can be attenuated by interference, path loss, static and mobile blockage, etc. The current 4G/LTE mobile access is much more volatile and unpredictable than WiFi and wireline accesses. While 5G mmWave transmission can deliver data rates over $1Gbps$, mmWave signals at higher frequency bands (20 -100 GHz) incur higher free-space path loss, blockage loss, and penetration loss~\cite{al2019millimetre}. The bandwidth variations experienced by users of the initial batch of commercial 5G deployments, both mmWave and Sub-6GHz, are much more dramatic than 4G/LTE~\cite{narayanan2020first,raca2020beyond}.

How to deliver a high level of user Quality-of-Experience under volatile mobile access conditions is a main challenge for mobile app developers and multimedia application service providers. For delay-tolerant applications, bandwidth variations can be ``absorbed" by sacrificing application-level latency. In the example of on-demand video streaming, a video buffer of tens of seconds is typically employed so that video can be streamed smoothly as long as the video rate matches the average bandwidth over ten seconds.  However, such long buffers are not possible for low-latency live video streaming and video conferencing. With low/no video buffer, the selected video rate has to closely track the instantaneous network bandwidth to avoid video freeze. To deliver high user QoE for such applications, the available bandwidth has to be accurately estimated in realtime to guide video rate adaption. Over-estimate will lead to video freeze, and under-estimate will lead to unnecessarily low video quality. Various realtime bandwidth prediction (for the next second) algorithms have been adopted by video streaming systems to guide video rate adaption~\cite{tian2012towards,winstein2013stochastic,jiang2014improving,yin2015control}. The emerging VR/AR/MR applications also hinge on high-rate and low-latency delivery of $360^\circ$ video/virtual objects over mobile connections to facilitate seamless integration of physical and virtual worlds and support user interactions. Realtime bandwidth prediction will again play an important role there. Another direction to cope with bandwidth variations is {\it multipath transmission}, such as MPTCP~\cite{mptcp}. Realtime bandwidth prediction can be used by multipath routing algorithms to {\it proactively} adjust the traffic split ratios among all the available paths.
In this paper, we study realtime prediction of available bandwidth and handoff in 4G/5G mobile networks. {\it Our first effort is to develop deep-learning based realtime bandwidth 
prediction models that generate predictions for the available bandwidth in the next few seconds based on the past bandwidth measurement as well as wireless channel and context information.} Specially, we focus on {\it fixed-route mobility} scenarios, covering routine daily commutes through public transportation and self-driving. We collect long consecutive 4G/LTE traces with a rich set of features in New York City MTA public transportation system. Through feature analysis,  we identify features with significance for predicting future bandwidth. We then demonstrate that Long Short Term Memory (LSTM) Recurrent Neural Networks~\cite{hochreiter1997long}, in particular TPA-LSTM~\cite{shih2019temporal}, can effectively mine the latent temporal patterns embedded in channel and context information for accurate {\it multivariate prediction}. Our LSTM-based prediction models consistently outperform the conventional univariate and multivariate prediction models in both 4G and 5G bandwidth traces.  

For handoff prediction, due to the limited initial 5G coverage, and the different deployment plan for 5G and 4G/LTE on urban and rural areas, 5G and 4G/LTE will co-exist in the long time. During an application session, a mobile device will likely switch back-forth between 5G and 4G access modes. Due to the vast disparity between QoS offered by 4G and 5G, it is of even greater importance to predict handoffs between the two access modes in realtime so that applications can make adjustments in advance to anticipate dramatic QoS changes resulted from handoffs. As the example of self-driving vehicle, 5G/4G brings different latency, which triggers totally disparate control strategies. {\it Our second effort is to predict  handoffs between 4G and 5G access modes based on realtime measurement of bandwidth and channel/context information}. We propose two versions of handoff prediction: in {\it binary prediction}, we predict whether the device will handoff from 4G to 5G or vice versa in the next second; in {\it continuous prediction}, we predict the probability/fraction of 5G access in a short future time window. We demonstrate that Gradient Boosting Machine (GBM)~\cite{friedman2001greedy} based classification and regression can achieve high accuracies in binary and continuous 4G/5G handoff predictions. 

The rest of the paper is organized as the following. The related work on realtime bandwidth prediction is reviewed in Section~\ref{sec:related}. We motivate and define the realtime bandwidth prediction problem for fixed-route mobility and present our LTE dataset in Section~\ref{sec:fixed-route}. TPA-LSTM prediction model is introduced in Section~\ref{sec:TPA-LSTM}, followed by prediction accuracy comparison with baselines. In Section~\ref{sec:5G}, we first present 5G bandwidth prediction results, then introduce the handoff prediction problem and present GBM-based classification and regression models for binary and continuous handoff prediction. The paper is concluded with future work in Section~\ref{sec:conclusion}.

\section{Related Work} 
\label{sec:related}
Realtime bandwidth prediction has been a challenging problem for the networking community. 

Authors of~\cite{jiang2014improving} and~\cite{yin2015control} used the Harmonic Mean of TCP throughput for downloading the previous chunks as the TCP downloading throughput prediction for the next chunk. A simple history-based TCP throughput estimation algorithm was proposed in~\cite{he2007predictability}. Authors of~\cite{kurdoglu2016real} used an adaptive filter,  Recursive Least Squared (RLS), to make realtime bandwidth prediction for the cellular scenario. For the conventional statistical and machine learning models, in~\cite{mirza2007machine}, the authors proposed that training a Support Vector Regression (SVR) model~\cite{smola2004tutorial} to estimate TCP throughput. In the context of DASH video streaming, in~\cite{tian2012towards}, authors adopted prediction algorithm in~\cite{he2007predictability} to guide the realtime chunk rate selection, and used a customized SVR model similar to~\cite{mirza2007machine} for DASH server selection. In the context of video conferencing, in~\cite{winstein2013stochastic}, the cellular link is modeled as a single-server queue driven by a doubly-stochastic service process, and future bandwidth prediction is generated by probabilistic inference based on the single-server queue model. Authors of~\cite{sun2016cs2p} used Hidden Markov Model (HMM) for bandwidth prediction. Authors of~\cite{yue2018linkforecast} proposed a Random Forest framework to make realtime LTE bandwidth prediction based on the context information. 
The conventional statistical or machine learning methods are based on short sequence history, and it is not easy to dig out the temporal patterns embedded in rich and complex information structures.  For deep leaning methods, in~\cite{ruan2019machine}, a Deep Neural Network (DNN) based method is applied for bandwidth burst prediction for Human to Machine (H2M) communication. \cite{wei2018trust}, ~\cite{mei2019realtime}, and~\cite{lee2020perceive} developed a Long Short Term Memory (LSTM)~\cite{gers1999learning} based method to estimate future bandwidth based on past bandwidth measurements. In~\cite{mei2020realtime}, authors discussed the feasibility of LSTM models on generalized application scenarios. For the multivariate time series we are facing, complex and non-linear inter-dependencies between variables at different time steps complicate the prediction task. Instead of the vanilla LSTM, we adopt the recently proposed Temporal Pattern Attention LSTM (TPA-LSTM)~\cite{shih2019temporal} for bandwidth prediction. It applies attention mechanisms to select the most relevant time steps and variables for prediction. For handoffs, \cite{ge2009history} and~\cite{kim2007mobility} studied handoff prediction between cells within the same access mode (LTE or 3G). We study handoffs between different access modes, specifically, between 4G and 5G.

\section{Realtime Bandwidth Prediction under Fixed-route Mobility}
\label{sec:fixed-route}
\subsection{Fixed-route Mobility}
\begin{figure}
\centering
\includegraphics[height= 1.7 in, width= 2.9 in]{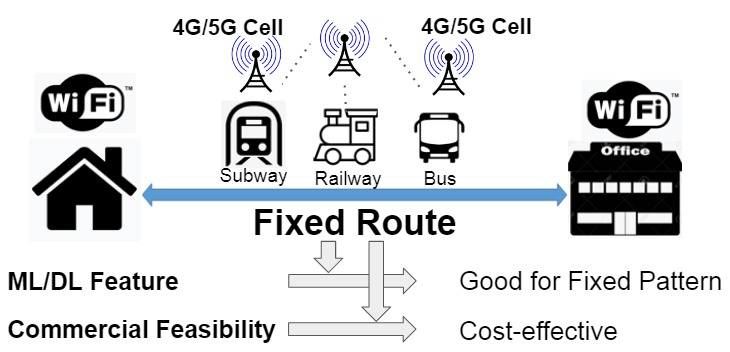}

\caption{BW Prediction for Fixed-route Mobility~\label{fig:SA4MBP}}
\end{figure}

Most people's daily network access patterns are rather predictable. He/She is either at home or office or on the way between the two, as shown in Figure~\ref{fig:SA4MBP}. 
At home and office, there is usually good WiFi coverage, leading to good network Quality of Service (QoS). For the commute between home and office, either driving or taking public transportation, the routes are also relatively fixed. In Metropolitan areas, commuters access the Internet through mobile 4G/5G connections. The mobile access bandwidth is inherently time-varying, especially with user mobility. It is, therefore, important to predict future bandwidth to deliver a high-level of application QoE to commuting users. We choose to focus on realtime bandwidth prediction for fixed-route mobility not only because it covers a wide range of daily commute scenarios, but also it enables a DNN model to learn the bandwidth variation regularity resulted from fixed-routes for accurate prediction.

\subsection{Realtime Bandwidth Prediction}
Let $b(t)$ be the bandwidth available for a user at time $t$. User mobile device periodically measures the quality of the mobile access link with a certain frequency, e.g., every $1$ second. 
It can obtain a discrete time series of $\{{X(t)}, t=1, 2, \cdots.\}$, where \(X(t) \in {\mathbb {R}}^n\) is a  vector of $n$ types of measured information, including $b(t)$ and other metrics about the connection. The realtime bandwidth prediction problem at time $t$ is to estimate the available bandwidth at some future time instant $t+\tau$ given all the collected measurements up to $t$: 
\begin{equation}
\hat b (t+\tau)=\mathbf f\left(\{{X(k)}, k=1, 2, \cdots, t\}\right), 
\end{equation}
where $\tau$ is the future horizon value. For prediction based on recent history, we use only \(\{X(t - w + 1), \cdots , X(t)\}\) to predict $b(t+\tau)$, where $w$ is the sliding window size. In {\it univariate bandwidth prediction}, we only use past bandwidth measurement to predict future bandwidth, namely 
\begin{equation}
\hat b (t+\tau)=\mathbf {f^{(u)}}\left(\{b(k), k=1, 2, \cdots, t\}\right).
\end{equation}
In a {\it multi-variate bandwidth prediction}, we use measured channel and context data in addition to the past bandwidth measurement for the prediction. 

For univariate bandwidth prediction, there are many methods to construct the prediction function $\mathbf {f^{(u)}} (\cdot)$, from simple history-repeat, i.e., $\hat b (t+\tau) = b(t)$, Exponential Weighted Moving Average (EWMA), $\hat b (t+1) = (1-\alpha) \hat b (t) + \alpha b(t)$, Harmonic Mean, $\hat b (t+\tau) = h/ {\sum_{k=0}^{h-1}  1/{b(t-k)}}$, etc., to more sophisticated signal processing approaches, such as Kalman filter~\cite{brown1992introduction} and Recursive Least Squares (RLS)~\cite{haykin2008adaptive}. In~\cite{kurdoglu2016real}, RLS is used for realtime bandwidth prediction. By assuming $\hat b (t+1)=\sum_{k=0}^{h-1} \omega (k) b(t-k)$, RLS can recursively find the coefficients $\mathbf \omega$ that minimize the weighted linear least squares cost function. \cite{kurdoglu2016real} showed that RLS achieves higher accuracy than other averaging and signal processing algorithms, such as Least Mean Square and EWMA etc. For multivariate bandwidth prediction, machine learning tools, such as Support Vector Machine~\cite{mirza2007machine} and  Random Forest~\cite{yue2018linkforecast}, have been proposed. In~\cite{mei2019realtime} and~\cite{wei2018trust}, they show that LSTM deep neural networks can achieve higher accuracy than the conventional bandwidth prediction methods.

\subsection{High-level Design and Rationale \label{chap:high_SA4MBP}}
For fixed-mobility, we propose a Deep-learning based smart agent for mobile bandwidth prediction.  Our framework consists of {\it data measurement}, {\it model training} and {\it model running} steps. Specifically, in the measurement step, for a fixed commute route, the agent repeatedly takes measurements on multiple trips from the start to the end.  Bandwidth and related metrics are recorded. In the model training step, one DNN model is trained offline for each commute route, using all the data collected from that route. 
For the running step, the agent picks the model trained for the current route to generate realtime bandwidth prediction at each time step based on recent measurement.  

 \textbf{Deep Learning for Prediction:} Deep Neural Networks (DNNs) have recently gained lots of momentum due to the dramatic increase in data volume and computing power. They have become the new state-of-art in specific fields, such as computer vision, speech recognition, and natural language processing, etc. What they have in common is {\it strong temporal and spatial patterns}. 

In particular, LSTM-based deep learning method has an unparalleled advantage over the conventional time series analysis tools due to its special recurrent kernel structure. {\it We explore the temporal patten of mobile bandwidth variations over fixed-routes using LSTM-based DNN.} 

 \textbf{Easy Adoption:} Users of public transportation systems have strong needs for realtime bandwidth prediction. For a user watching online video, the video player can adapt the quality of the video to be downloaded based on the realtime downlink bandwidth prediction. For a user in a video conference, the conferencing app can dynamically change the resolution and frame rate of the video to be coded and uploaded to other users in the conference, based on realtime uplink bandwidth prediction. Under our proposed smart agent, using off-the-shelf software and hardware, any ordinary smartphone can easily take the measurements. In practice, measurement can be done by network optimization team, by crowd-sourcing, or even by transportation company. The offline trained prediction models can run in realtime on local phone or on edge to improve user QoE in various mobile apps.

\begin{figure*}

\centering
\subfloat[Subway 7 Train \label{fig:route t7}] {
       \includegraphics[height=1.2in,width=1.2in]{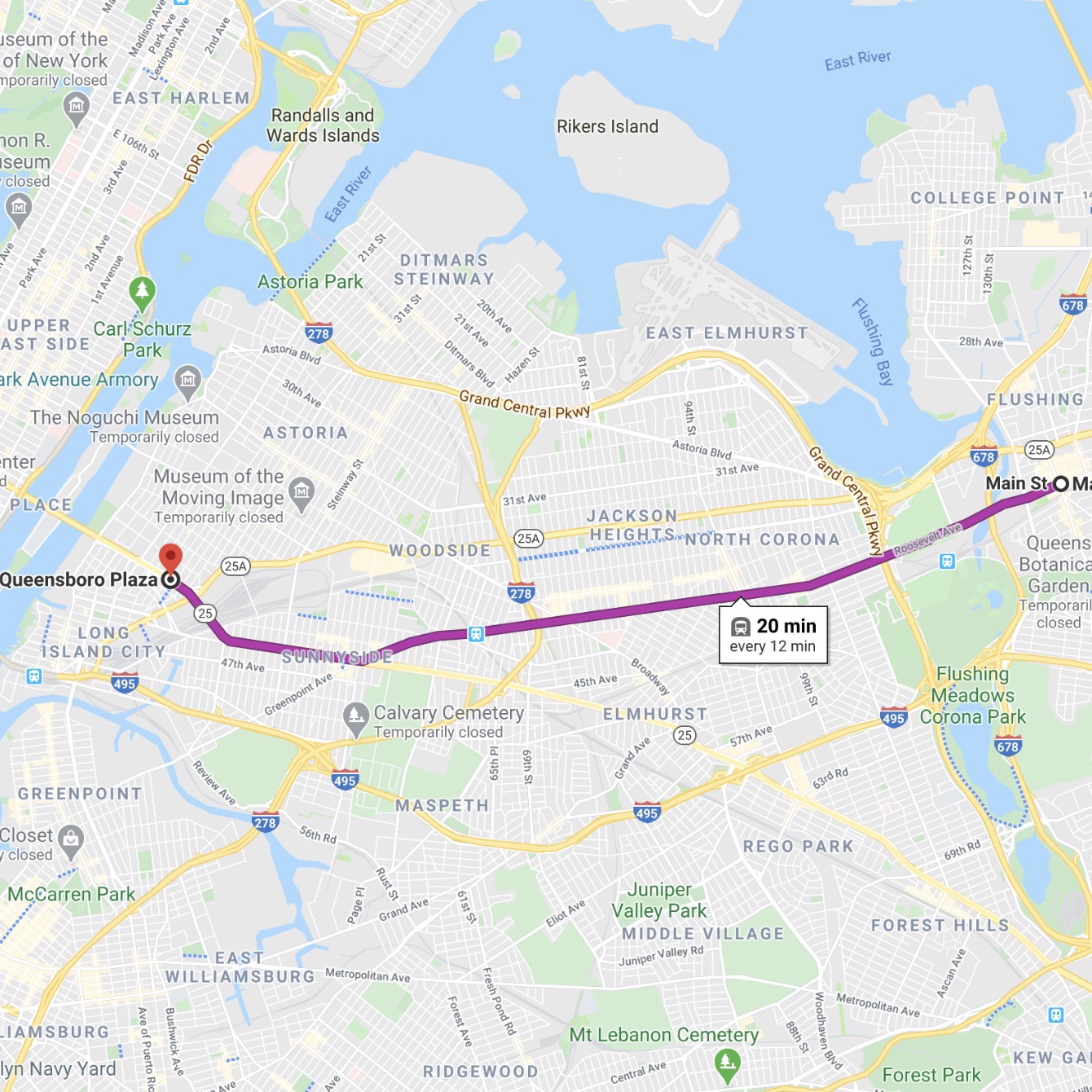}
     }
     \hfill
 \subfloat[Bus M15 \label{fig:route m15}] {
        \includegraphics[height=1.2in,width=1.2in]{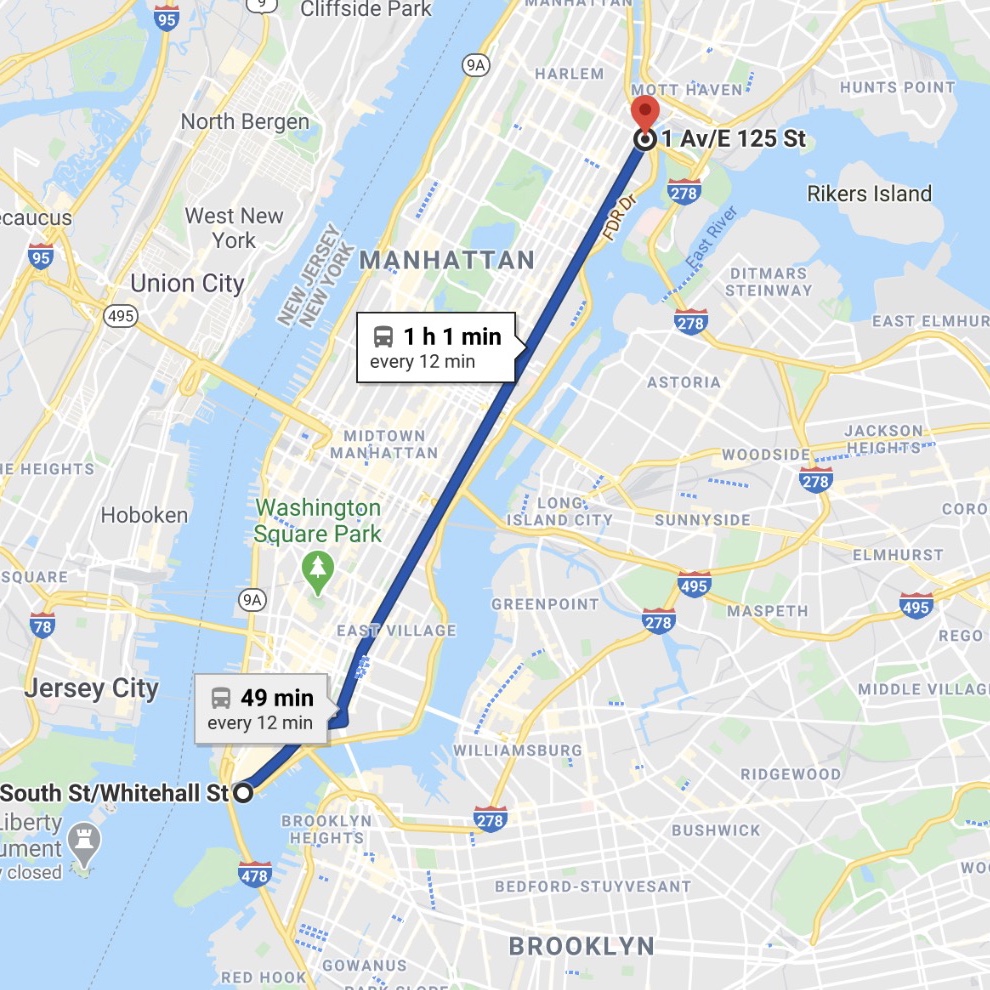}
     }
     \hfill
 \subfloat[Bus B16 \label{fig:route b16}] {
       \includegraphics[height=1.2in,width=1.2in]{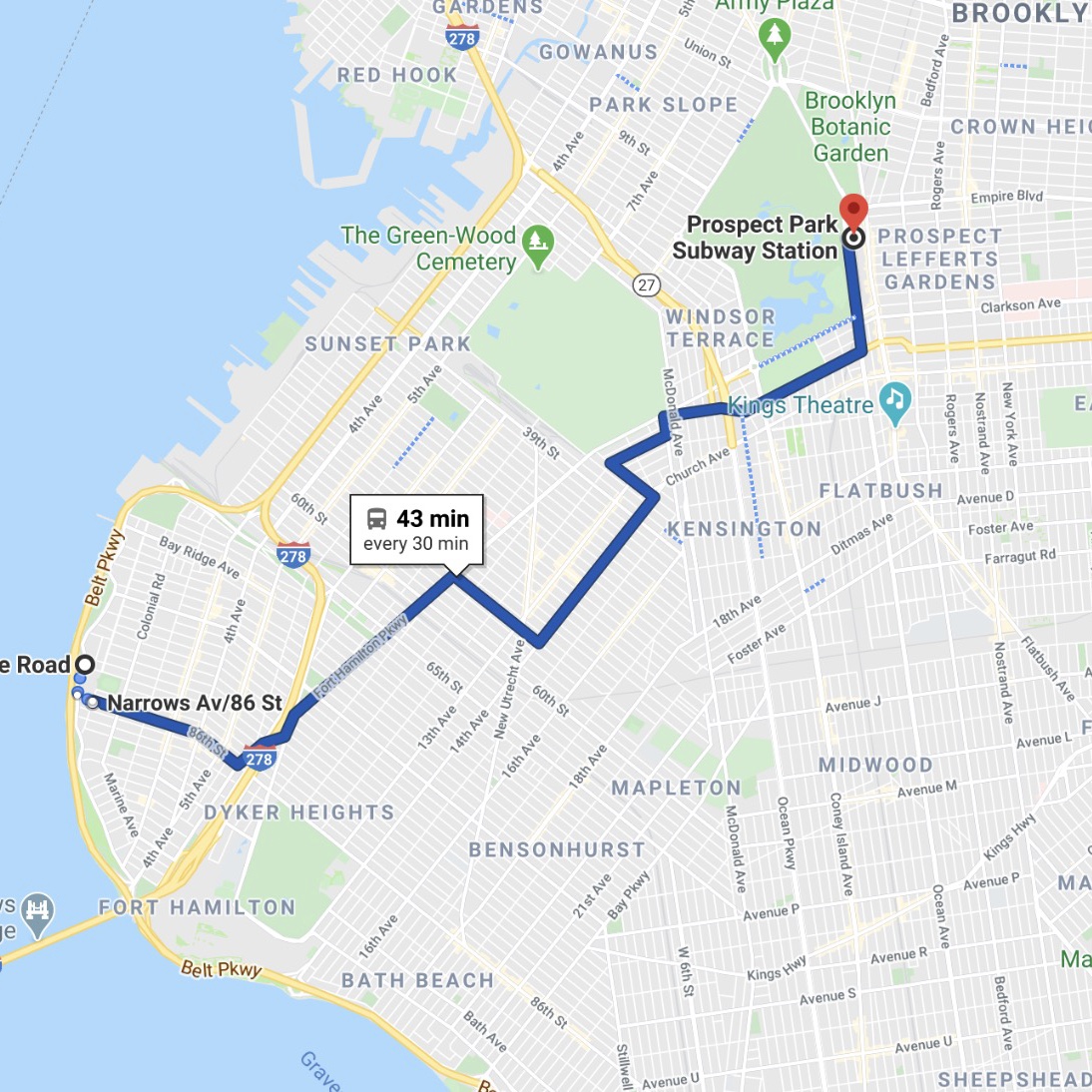}
     }
     \hfill
\subfloat[Bus B61 \label{fig:route b61}] {
        \includegraphics[height=1.2in,width=1.2in]{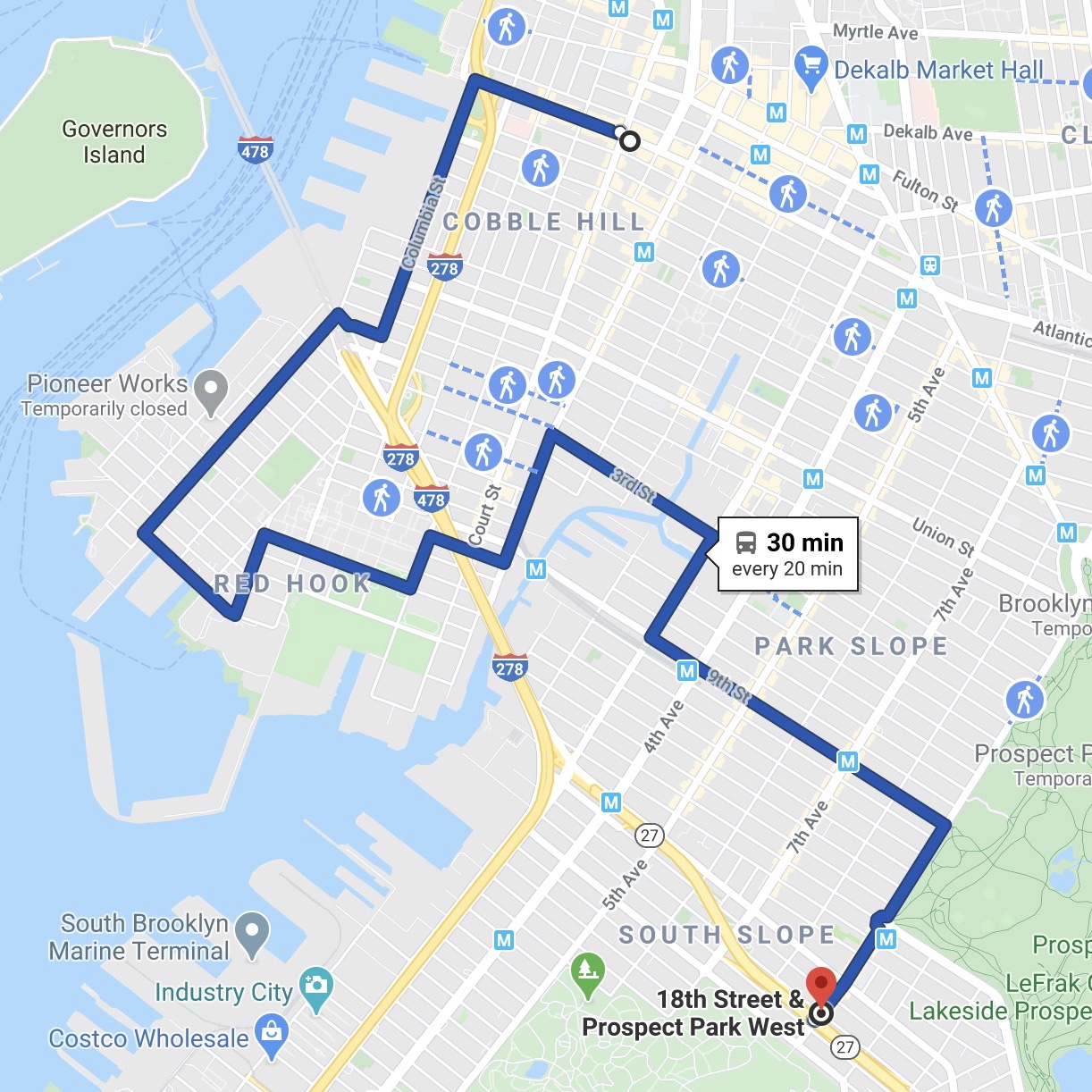}
     }
     \hfill
\subfloat[Bus B62 \label{fig:route b62}] {
       \includegraphics[height=1.2in,width=1.2in]{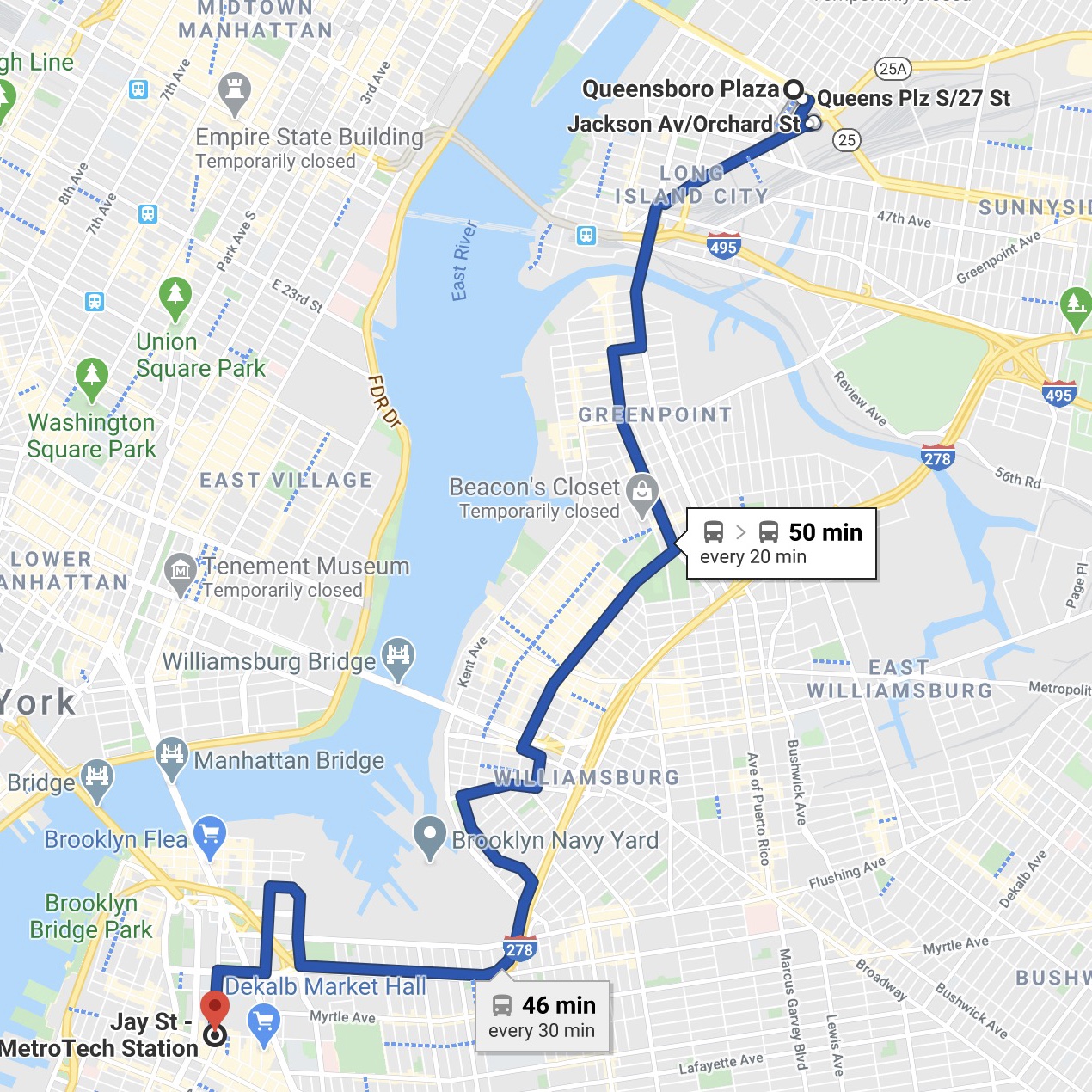}
     }
     \hfill

\caption{Sample LTE Measurement Routes of Public Transportation System of New York City~\label{fig:MTARoutes}}  

\end{figure*}

\begin{figure}
\subfloat[Subway 7 Train \label{fig:bw t7}] {
       \includegraphics[height=1.4in,width=1.5in]{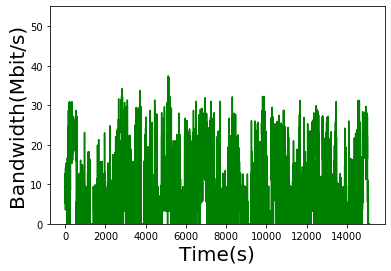}
     }
 \subfloat[Bus B62  \label{fig:bw b62}] {
        \includegraphics[height=1.4in,width=1.5in]{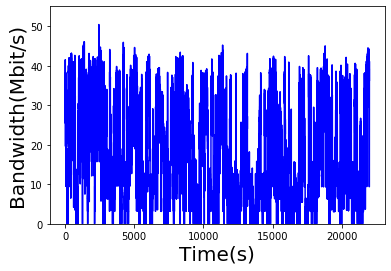}
     }
  \caption{Sample LTE Bandwidth Traces from Subway \#7 and Bus \#B62 \label{fig:samples}}
\end{figure}

\subsection{LTE Dataset in NYC}
Towards our goals, we conducted a measurement study on the public transportation system of NYC. Different from other LTE mobile bandwidth datasets, our dataset is multivariate and focused on fixed public transportation routes. We pick five bus/subway routes of the NYC MTA system, as illustrated in Figure~\ref{fig:route t7} to~\ref{fig:route b62}. The LTE data is collected from Nov, 2019 to the end of Jan, 2020. For each route, we collected long uninterrupted traces by taking around eight trips from one end to the other in both directions, with the duration of each trip to be more than 30 minutes. The total duration of our LTE dataset is around $30$ hours. 
 We measured bandwidth, channel, and context related information using $Net Monitor Pro$, a mobile network monitoring tool designed for Android devices. We installed this app on a Google Pixel 1 phone with unlimited 4G LTE Data Plan. To record the bandwidth, we run $iPerf$ to download data from our lab server located on our NYC campus, and record TCP download throughput every $1,000$ millisecond. The bandwidth is logged on the mobile phone with timestamps, so we can use timestamps to match the channel and context information logged by $Net Monitor Pro$. Figure~\ref{fig:bw t7} and~\ref{fig:bw b62} visualize two collected bandwidth traces, one from a subway, one from a bus. Table~\ref{tbl:stat 4G} presents the bandwidth statistics of the collected traces. 

\begin{table*}
\centering  

\caption{Statistics of NYC Public Transportation Bandwidth Traces (Mbps)~\label{tbl:stat 4G}}

\begin{tabular}{lccccc}  
\hline
  &7 Train	&Bus B16	&Bus B61	&Bus B62	&Bus M15\\ \hline  
Average  &8.67	&13.71	&17.80	&18.34	&20.63 \\\hline
Median &6.85	&12.80	&16.00	&15.80	& 19.10\\\hline
Max &37.40	&45.00	&47.40	& 50.40	&44.30\\\hline
Min &0	&0	&0	&0	&0\\\hline
Std &8.29	&9.55	&11.69	&12.16	&9.53\\\hline
Length (s) &15,116 &22,277 &21,174 &22,000 &23,103\\\hline

\end{tabular}

\end{table*}

\subsection{LTE Feature Analysis~\label{Sec:LTE_Features_ANAL}}
Other than the bandwidth, we recorded a total of 52 mobile channel and context related features. Out of all the features, we conduct feature selection by calculating the cross-correlation between each feature and bandwidth. Only eight features are selected as the input for the prediction models, as illustrated in Table~\ref{tbl:features}.  We further analyzed the importance of each selected feature for bandwidth prediction using Random Forest~\cite{liaw2002classification}. We use all the eight features at $t-1$ as input of a Random Forest model, the output is the bandwidth at $t$. Table~\ref{tbl:importance 4G} shows the importance weights of all features on all traces.

\begin{table}\caption{Selected Features}
\begin{center}
\small
{\begin{tabular}{lp{0.6\linewidth} }

\hline
Feature & Information Captured\\ 
\hline
\texttt{Bandwidth(BW)}&  download throughput in Mbps\\ 
\texttt{LTE-neighbors} & number of LTE cells the device can switch to \\
\texttt{RSSI}   &  power level of received signal\\
\texttt{RSRQ} & quality of received signal \\
\texttt{Echng(Ech)} & whether ENodeB has changed from previous second\\
\texttt{TA}  & time advance needed to reach the ENodeB\\
\texttt{Speed} & moving speed of device\\
\texttt{Band} & frequency band of signal\\ 
\hline
\end{tabular}

\label{tbl:features}
}
\end{center}
\end{table}

\begin{table*}
\centering  
\caption{Feature Importance on LTE Dataset~\label{tbl:importance 4G}}

\begin{tabular}{lcccccccc}  
\hline
  &\texttt{Bandwidth} &\texttt{Band}  &\texttt{RSSI} &\texttt{RSRQ} &\texttt{TA}  &\texttt{LTE\_neighbors}&\texttt{Speed}&\texttt{Echng}\\ \hline  
7 Train&0.5805 &0.1105 &0.1189	&0.0601	&0.0393	&0.0540 &0.0338 &0.0028 \\\hline
Bus B16 &0.6698 &0.1382 &0.0612 &0.0283 &0.0585 &0.0195 &0.0230 &0.0014 \\\hline
Bus B61 &0.5170  &0.1253 &0.1890	&0.0634	&0.0660	&0.0192 &0.0184 &0.0017 \\\hline
Bus B62 &0.5392  &0.2310 &0.1212	&0.0294	&0.0491	&0.0119 &0.0173 &0.0009 \\\hline
Bus M15 &0.6062 &0.2383 &0.0497	&0.0281	&0.0178	&0.0269 &0.0319 &0.0010 \\\hline
\end{tabular}

\end{table*}

As expected, \texttt{Bandwidth} at $b(t-1)$ has the highest prediction weight on $b(t)$. \texttt{Band} has the second highest weight. This is because high speed LTE data transmission is provided in frequency bands like Band 1900 and 2100, while relatively low speed transmission is provided in others, like Band 700.
Bandwidth tends to be high when the signal is transmitted in an ideal band, and becomes low when switched to a non-ideal band. The third most important feature is \texttt{RSSI}, which indicates whether the signal power between the base station and the mobile device is strong or not. Meanwhile, feature  \texttt{RSRQ}'s weight is not as high, since it is calculated from \texttt{RSSI}. The importance weights of the remaining features are not very high. But this does not mean they are not important for bandwidth prediction. It is only that when all eight features are used together, their prediction power is dominated by other more powerful features, such as \texttt{Bandwidth}, \texttt{Band}, and \texttt{RSSI}. But they still might provide complementary information in certain scenarios. For example, \texttt{Echng} indicates handoff of ENodeB. Since our device keeps moving, the handoffs occur frequently. Due to handoff, there is a short period of time when the device receives no service from any ENodeB and bandwidth dips to zero. We recorded the ID of the ENodeB that our device is connected to. Whenever the ENodeB ID changes, we consider a handoff happens. Similarly, \texttt{Speed} should be an important feature for mobile bandwidth prediction. Both the signal quality and handoff frequency are highly dependent on device moving speed. But its importance weight calculated by Random Forest is not high. This suggests that its impact on bandwidth is indirectly carried by  \texttt{BAND/RSSI} and \texttt{Echng}.  {\it We feed all the eight selected features to our DNN models, which are expected to exploit the complementary prediction power of all features (better than Random Forest) for more accurate prediction.}

\section{TPA-LSTM based Bandwidth Prediction}
\label{sec:TPA-LSTM}
   \subsection{Introduction to TPA-LSTM}
\begin{figure*}

\captionsetup[subfigure]{justification=centering}

\subfloat[LSTM Unit~\label{fig:LSTM-Unit}]{
  \includegraphics[height=1.44in,width=0.3\linewidth]{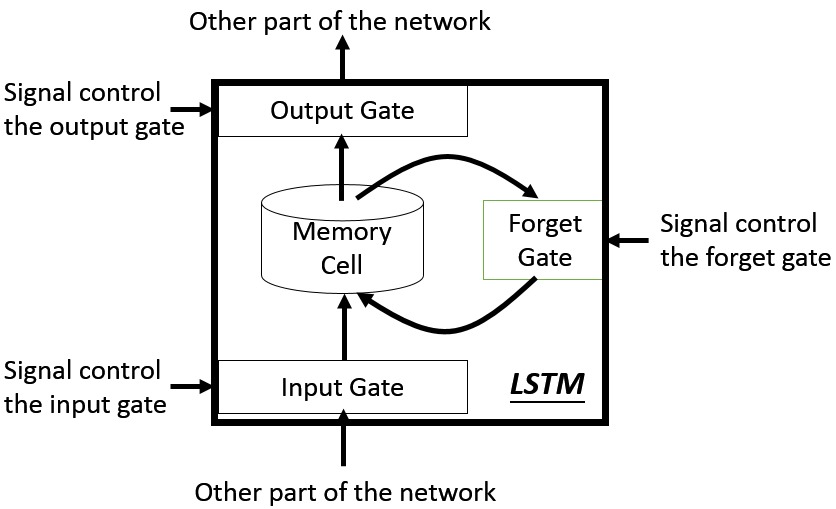}}
 \subfloat[TPA-LSTM~\label{fig:TPA-LSTM}]{
  \includegraphics[height=1.44in, width=0.6\linewidth]{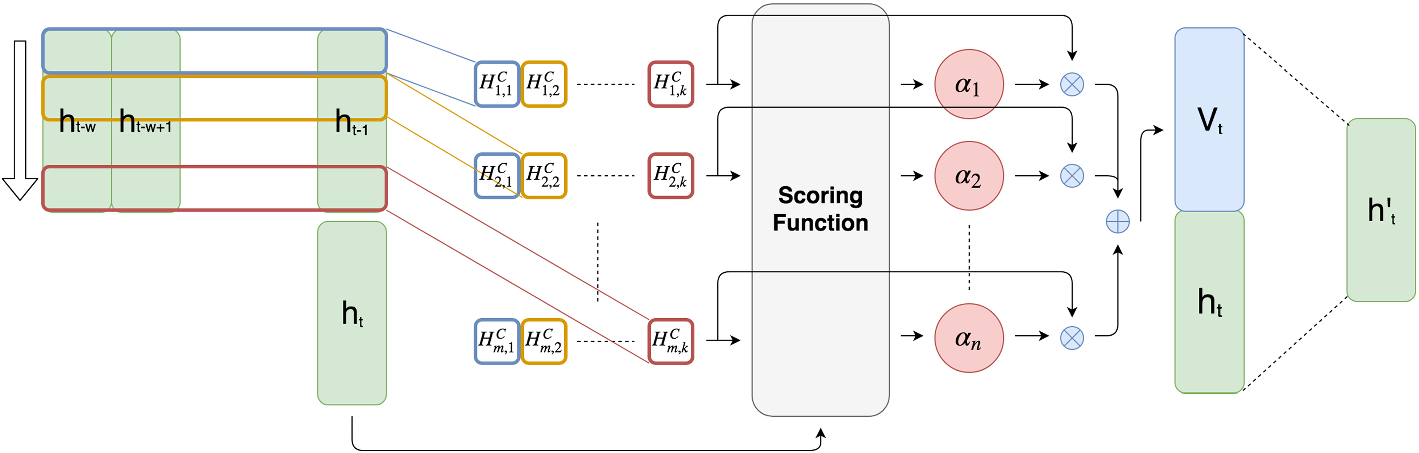}}

  \caption{DNN Architecture: a) Internal Structure of LSTM Unit; b) TPA-LSTM Network~\label{fig:DNN-Arch}}
\end{figure*}

We pick Temporal Pattern Attention Long Short-Term Memory (TPA-LSTM)~\cite{shih2019temporal} as our core DNN prediction model. TPA-LSTM extends the vanilla LSTM~\cite{hochreiter1997long} with the Attention Mechanism~\cite{wang2016attention}. Figure~\ref{fig:LSTM-Unit} shows the internal structure of the vanilla LSTM unit. LSTM shows great performance in time series analysis due to its special internal memory cell, forget gate, input gate, and output gate. Due to recurrent update, LSTM can keep  ``long memory" of a time series. Through training, LSTM can learn which part of a time series is important and which is not for predicting the output. Meanwhile, the conventional Attention Mechanism looks back information from the previous time steps, uses 
the relevance to improve the prediction accuracy. But it is difficult to deal with long time sequences.
TPA-LSTM~\cite{shih2019temporal} combines the merits of LSTM and Attention Mechanism.  It uses a set of Convolution Neural Network (CNN) filters to extract time-invariant temporal patterns. It makes the dimension of attention to be feature-wise, so that it can learn the inter-dependencies among multiple features over a long history time window. Figure~\ref{fig:TPA-LSTM} illustrates the architecture of TPA-LSTM. Let $h_t$ be the LSTM hidden state at time $t$. Instead of using $h_t$ to generate prediction for $t+1$, TPA-LSTM learns the ``importance" of the hidden states $\{h_{t-1}, \cdots, h_{t-w}\}$ of the previous $w$ steps. Specifically, $k$ 1-D CNN filters with length $w$ are applied to $\{h_{t-1}, \cdots, h_{t-w}\}$, as shown in Figure~\ref{fig:TPA-LSTM}. Each CNN filter makes convolution over hidden feature values. All filters produce a new matrix $H^C$. The attention part calculates the attention (importance) weights of all the convoluted hidden features. The scoring function calculates a weight for each row of $H^C$ by comparing it with the current hidden state $h_t$. The rows of $H^C$ is then weighted summed by their corresponding weights to get a new vector $V_t$, which is concatenated with $h_t$ to generate an updated hidden state $h^\prime _t$ for the final prediction~\cite{shih2019temporal}. The CNN filters and attention calculation enhance the capability of vanilla LSTM to mine periodic temporal patterns in time series. Through experiments on Multivariate Time Series datasets, such as currency exchange rate among several countries and electricity among multiple clients, it has been demonstrated that TPA-LSTM can achieve higher accuracy than LSTM in multivariate time series prediction, even when the periodic pattern is weak~\cite{shih2019temporal}. For our mobile bandwidth prediction problem, in some cases, the future bandwidth may be more dependent on history data farther back than the most recent history. TPA-LSTM,  using its CNN, can look back further to dig out the inter-dependencies among multiple variables that are multiple time steps apart. In addition, TPA-LSTM perfectly suits the fixed-route mobility scenarios under study. As mentioned in Section~\ref{chap:high_SA4MBP}, {\it for repeated long trips along the same route, periodic patterns on bandwidth, and the related channel features are expected within a single trip and cross multiple trips.  TPA-LSTM can effectively mine those patterns for more accurate prediction.} 

\subsection{Prediction Performance\label{chap:Pred_perf}}

\subsubsection{Methods for Comparison}
\

\noindent Here we make a comparison between TPA-LSTM method and other baselines univariate and multivariate methods:

\begin{itemize}
\item RLS: Recursive Least Square adaptive algorithm~\cite{kurdoglu2016real}

\item RF: Random Forest~\cite{yue2018linkforecast} 

\item LSTM: Vanilla Long Short-Term Memory~\cite{wei2018trust}
\end{itemize}
 Among them, RLS is for univariate with previous bandwidth measurement as its only input feature. The rest of the methods are for multivariate bandwidth prediction, with all the eight features as the input.

\subsubsection{Model Settings}
\
\noindent For neural network methods, the network structure is: 3 layers, and every layer has 32 units with dropout 0.2. We divide our LTE dataset into a training set, validation/development set, and test set, according to ratios of $60\%:10\%:30\%$. We use Adam~\cite{kingma2014adam} as optimizer with the default learning rate of $0.001$. The loss function is Mean Square Error (MSE) of the predicted bandwidth. For features unique to TPA-LSTM method~\cite{shih2019temporal}, we set the CNN filter number to 32.
For Random Forest, we use the same model setting as~\cite{yue2018linkforecast}. We set the criterion as Mean Square Error (MSE). The \texttt{max-features} is set to be Square Root (SQRT). To obtain good performance, the number of decision trees is set to be $1,200$. The max-depth of the trees is set to be $20$. The minimum number of samples to be split is $10$. The minimum samples kept in one leaf is set to be $2$.

\subsubsection{Performance Metrics}
\
\noindent We use Root Mean Square Error (RMSE) and Mean Absolute Error (MAE) as the main metrics for prediction errors, and use  Pearson Correlation (CORR), ranging from $-1$ to $1$, as the reference metric for sequence correlation between the prediction and the ground-truth.

\begin{table}

\caption{Evaluation on Bus M15
  ~\label{tbl:ReM15}}
\begin{center}
\begin{tabular}{lllll}\hline
M15         &   Mean: & 22.7147       & Std: & 9.2526   \\\hline
Horizon = 1 & RLS        &  RF & LSTM      & TPA-LSTM \\\hline
RMSE        & 4.7040     & 4.4912        & 4.1899    & {\bf4.0038}   \\
MAE         & 3.3713     & 3.3682        & 3.1052    & {\bf2.9043}   \\
CORR        & 0.8647     & 0.8804        & 0.8939    & {\bf0.9025}   \\\hline
            &            &               &           &          \\\hline
Horizon = 2 & RLS        &  RF & LSTM      & TPA-LSTM \\\hline
RMSE        & 5.1766     & 4.7689        & 4.6514    & {\bf4.6102 }  \\
MAE         & 3.6173     & 3.4236        & 3.2973    & {\bf3.2362 } \\
CORR        & 0.8357     & 0.8601        & 0.8663    & {\bf0.8671  }\\\hline
            &            &               &           &          \\\hline
Horizon = 3 & RLS        &  RF & LSTM      & TPA-LSTM \\\hline
RMSE        & 5.5763     & 5.2344        & 5.1524    & {\bf5.0779}   \\
MAE         & 3.8503     & 3.7594        & 3.6317    & {\bf3.5488}   \\
CORR        & 0.8087     & 0.8288        & 0.8309    & {\bf0.8362}  \\\hline
\end{tabular}
\end{center}
\end{table}

\begin{table}

\caption{Evaluation on Subway Train 7
  ~\label{tbl:ReT7}}
\begin{center}
\begin{tabular}{lllll}\hline
Train7      &   Mean: & 8.3430        & Std: & 8.0403   \\\hline
Horizon = 1 & RLS        &  RF & LSTM      & TPA-LSTM \\\hline
RMSE        & 4.8929     & 4.6958        & 4.3964    & {\bf4.3483}   \\
MAE         & 3.4632     & 3.5091        & 3.2936    & {\bf3.1941}   \\
CORR        & 0.8003     & 0.8120        & 0.8405    & {\bf0.8414}   \\\hline
            &            &               &           &          \\\hline
Horizon = 2 & RLS        &  RF & LSTM      & TPA-LSTM \\\hline
RMSE        & 5.0092     & 4.8089        & 4.6065    & {\bf4.5884}   \\
MAE         & 3.4660     & 3.5488        & 3.3622    & {\bf3.2710}   \\
CORR        & 0.7901     & 0.8025        & 0.8216    & {\bf0.8222}   \\\hline
            &            &               &           &          \\\hline
Horizon = 3 & RLS        &  RF & LSTM      & TPA-LSTM \\\hline
RMSE        & 5.3253     & 4.9537        & 4.8869    & {\bf4.8781}   \\
MAE         & 3.7218     & 3.6556        & 3.5845    & {\bf3.4617}   \\
CORR        & 0.7610     & 0.7900        & 0.7953    & {\bf0.7963}  \\\hline
\end{tabular}
\end{center}
\end{table}

\begin{table*}%

\centering
\caption{RMSE for B61, B62, and B16 Bus Traces}

\label{tbl:RMSE_B61B62B16}

{

\begin{tabular}{||c||c|c|c|c||c|c|c|c||c|c|c|c||}

\hline

\bfseries  & \multicolumn{4}{|c||}{\bfseries Horizon = 1} & \multicolumn{4}{|c||}{\bfseries Horizon = 2} & \multicolumn{4}{|c||}{\bfseries Horizon = 3} \\ \cline{2-13}

 \bfseries RMSE & \bfseries RLS & \bfseries RF & \bfseries LSTM   &  \bfseries TPA & \bfseries RLS  &  \bfseries RF     &  \bfseries LSTM   & \bfseries TPA & \bfseries RLS         & \bfseries RF     &\bfseries LSTM   &\bfseries TPA\\ 

\hline

B61  & 4.3403      & 4.8865 & 4.2137 & {\bf 4.0452}   & 4.7124      & 4.8845 & 4.5162 & {\bf 4.4607}   & 5.1202      & 5.2650 & 4.9645 & {\bf 4.8987}   \\
B62  & 4.8097      & 4.6408 & 4.4797 & {\bf 4.2138}   & 5.3032      & 5.0046 & 4.8975 & {\bf 4.8286}   & 5.7675      & 5.4324 & 5.4020 & {\bf 5.3250}   \\
B16  & 3.7693      & 5.4915 & 3.6762 & {\bf 3.5362}   & 3.9959      & 4.7346 & 3.8984 & {\bf 3.7348}   & 4.2284      & 5.0466 & 4.0465 & {\bf 3.9697}   \\ \hline

\end{tabular}
}

\end{table*}

\subsection{Results and Analysis}

We compare the performance of all the models at different prediction horizons, $\tau=1,2,3$ seconds. For the history window size, we set $W$ to $5$ for RLS, LSTM, and TPA-LSTM.  For RF, according to the conclusion from~\cite{yue2018linkforecast}, a too large $W$ would decrease the accuracy, so we set $W = \tau$.
 
Table~\ref{tbl:ReM15} to Table~\ref{tbl:ReT7} show the detailed evaluation result on two representative dataset traces Bus M15 and Subway Train 7. Table~\ref{tbl:RMSE_B61B62B16} shows the RMSE from B61, B62, and B16 for Horizon equals 1 to 3. The unit over all datasets is Mbps. Among those tables, bold fonts represent the best one for each metric. We can find that TPA-LSTM is the best method on RMSE, MAE, and CORR almost over all the datasets and all the prediction horizons. Taking $\tau= 1$ as an example, for RMSE, TPA-LSTM is on average $11.7\%$  better than the other methods; the improvement over the second-best method is $3.6\%$. For MAE, TPA-LSTM is on average $15.28\%$ better than other methods, the improvement over the second-best is $5.8\%$. It shows that TPA-LSTM fits our bandwidth prediction problem well. Also, for the horizon value, as the horizon becomes longer, the prediction performance of each algorithm gets worse. The decreasing trend is also reflected by CORR values. E.g., in Table~\ref{tbl:ReT7}, CORR of TPA-LSTM decreases  from 0.8414 to 0.7963 when horizon increases from 1 to 3. 

In addition, we found that the relative performance order is RLS$<$RF$<$LSTM$<$TPA-LSTM in general. As a univariate algorithm, RLS is not as good as the other multivariate algorithms.  Even though RLS is already a good adaptive filter for univariate bandwidth prediction~\cite{kurdoglu2016real}, it cannot utilize other useful channel and context information, so it hardly can reach the performance of the other multivariate algorithms. Random Forest is multivariate, and utilizes the channel and context information through feature pattern mining. However, for multivariate time series, it is also very important to mine temporal patterns in the long-term time domain. Random Forest does not have enough mining capacity for the long-term temporal patterns. In~\cite{yue2018linkforecast}, it was shown that longer history window size not only cannot help prediction, but could decrease the performance. LSTM-based algorithms are designed to mine both long and short-term temporal patterns.  This explains why LSTM-based algorithms are better than Random Forest even though they use the same input features. Taking Train7 in Table~\ref{tbl:ReT7} as an example, at horizon $1$, RMSE of RF is $4.6958$, however LSTM and TPA-LSTM are $4.3964$ and $4.3483$, repectively. The gap is large. For MAE,  the gap is also clear. MAE of RF is $3.3591$, but for LSTM and TPA-LSTM, MAE are $3.2936$ and $3.1941$ respectively. LSTM and TPA-LSTM have similar architecture: recurrent neural network between input and output, and use various gates to control input and output. That is why the performance gap between LSTM and TPA-LSTM is smaller than the gaps to the others. The small gain of TPA-LSTM comes from its attention mechanism, which allows it to work with longer time windows and a wider range of features at each time step.

\section{Bandwidth and Handoff Prediction in 5G Networks}
\label{sec:5G}
\begin{figure}
\centering
\includegraphics[height=2in]{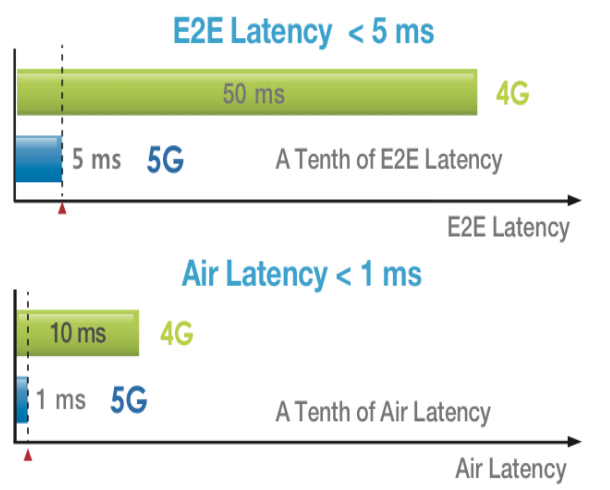}
\caption{Target Latency in 5G vs 4G~\cite{5g4glatency} ~\label{fig:5g_latency}}
\end{figure}

The fifth generation (5G) mobile networks have started to be deployed for commercial use world-wide since 2019. It will become more and more prevalent in the near future. 5G not only promises higher bandwidth throughput but also lower latency than 4G LTE. Figure~\ref{fig:5g_latency} from~\cite{5g4glatency} visualizes the $10x$ reduction of the target end-to-end and air latency from 4G to 5G.  
At the same time, 5G PHY layer operates at higher frequency bands, e.g., millimeter Wave (24-100GHz), which are more vulnerable to higher free space path loss, blockage loss, and penetration loss~\cite{al2019millimetre}. This poses a significant challenge to deliver stable 5G mobile access.

\subsection{Realtime 5G Bandwidth Prediction}
\subsubsection{5G Dataset}
We were planning to extend our measurement campaign to commercial 5G deployment in NYC after finishing our LTE measurement. However, due to the unexpected COVID-19 pandemic, we could not go out for taking measurements in the hardest hit city in the world. We had to turn to 5G datasets collected by other groups before the pandemic. We got to know two recent public datasets; one is from the University of Minnesota (UMN), USA~\cite{narayanan2020first}, the other one is from the University College Cork (UCC), Ireland~\cite{raca2020beyond}. The UMN trace was collected in several metropolitan areas in USA. However, their traces are relatively short, e.g. around 300 seconds per mobility trace, and the channel information and context information were not published. On the other hand, the UCC 5G traces were collected in Cork, Ireland with uplink/downlink bandwidth, as well as rich channel and context information. Even though the traces were not collected on the public transportation system, we picked the car driving traces along fixed-routes (based on analyzing GPS location information) and focus on download bandwidth prediction to make them comparable to our own LTE traces. 

Since 5G networks do not have complete coverage yet, the current practice is to fall back to 4G/LTE whenever a UE moves outside of 5G coverage. 

In the driving traces, the mobile access mode alternates between 5G and 4G. Table~\ref{tbl:stat 5g} shows the full statistics of bandwidth on the 5G dataset. The first row is the overall statistics for mixed 5G/4G access modes. The total length is $18,043$ seconds. The average bandwidth in Mpbs is $39.78$. The median is $12.68$. The highest is $532.91$. The lowest is 0. The standard deviation is as high as $66.73$, even twice as the average value! The second row is for 5G access mode only, and the third row is for 4G access mode only. It is obvious that while the 5G access mode has higher average bandwidth than 4G access mode, it also has  higher variance than 4G as expected. It poses a significant challenge for accurate bandwidth prediction, especially when the access mode switches back-forth between 5G and 4G. 
\begin{table*}
\centering  
\caption{Statistics of UCC 5G Driving  Dataset (Mbps)~\label{tbl:stat 5g}}
\begin{tabular}{lcccccc}  
\hline
  &Average	&Median	&Max	&Min	&Std & Duration (s)\\ \hline  
5G/4G &39.78	&12.68	&532.91	&0	&66.73  &18,043 \\\hline
5G-only &57.35	&19.887	&532.91	&0	&78.95  &10,837 \\\hline
4G-only &13.36	&8.33	&372.53	&0	&24.77  &7,210 \\\hline
\end{tabular}
\end{table*}

\subsubsection{5G Features Analysis}
\
\noindent There are totally $26$ features recorded in the UCC dataset. Similar to Section~\ref{Sec:LTE_Features_ANAL}, we selected $12$ features for 5G bandwidth prediction using the Random Forest for importance analysis, as shown in Table~\ref{tbl:5g-features}. \texttt{RSSI} and \texttt{RSRQ} have similar meaning as in 4G. \texttt{NRxSRP} and 
\texttt{NRxSRQ} are signal quality in neighboring cells (NR stands for Neighbor). \texttt{Cell-handoff} is similar to \texttt{Echng} for LTE, indicating handoff between cells. The feature importance for predicting the next-second download bandwidth is reported in Table~\ref{tbl:importance 5G}. 
 We can find that the feature with the highest importance is \texttt{DL}, with the importance of $0.585$. The second highest is \texttt{UL} with importance of $0.17$. The importance of other features are less than $0.10$, but they are still taken into account for bandwidth prediction because they have complementary channel and context information for bandwidth prediction.
 
\begin{table}\caption{Selected 5G Features}
\centering
\small
{\begin{tabular}{l c p{0.5\linewidth} }

\hline
Feature && Information Captured\\ 
\hline
\texttt{DL}& & downlink throughput in Mbps\\ 
\texttt{UL} &&  uplink throughput in Mbps\\
\texttt{RSSI}   &&  power level of received signal\\
\texttt{RSRQ} && quality of received signal \\
\texttt{RSRP} && reference signal receive power \\
\texttt{NRxSRP} && RSRP in neighbor cell\\
\texttt{NRxSRQ} && RSRQ in neighbor cell\\
\texttt{SNR}   &&  ratio of signal power to the noise power (in db)\\
\texttt{CQI} && channel quality indicator\\
\texttt{NetworkMode} && current access mode (5G/LTE)\\
\texttt{Cell-handoff}  && indicator for horizontal handoff between cells of the same mode\\
\texttt{Speed} && moving speed of device\\
\hline
\end{tabular}

\label{tbl:5g-features}
}
\end{table}

\begin{table*}
\centering

\caption{Feature Importance on 5G Dataset in percentage(\%)~\label{tbl:importance 5G}}
\begin{tabular}{lcccccccccccc}  
\hline
DL    & UL    & NRxSRP & RSRP   & NetworkMode & RSSI   & NRxSRQ & Speed  & SNR    & RSRQ   & CQI    & Cell-handoff \\ \hline
58.50 & 17.40 & 5.04 & 3.95 & 3.81  & 2.54 & 2.41 & 1.98 & 1.97 & 1.21 & 1.14 & 0.05    \\ \hline
\end{tabular}
\end{table*}

\subsubsection{Prediction Results}
\noindent
Table~\ref{tbl:Re5G} shows the bandwidth prediction accuracy on the 5G driving trace. The models and parameter settings are the same as in Section~\ref{chap:Pred_perf}. We can find that on the 5G dataset, TPA-LSTM is still better than the other prediction models. We can also find that the accuracies for all methods are universally worse than their accuracies in our LTE datasets. This is expected because the 5G signal is more dynamic: the mean of the testset is around $27.579$, however, the standard variance is as high as $51.276$, which is totally different than the five LTE datasets. For Horizon = $3$ in Table~\ref{tbl:Re5G}, Random Forest is slightly better than TPA-LSTM in terms of RMSE and CORR, while TPA-LSTM is better in terms of MAE. The large variance of 5G bandwidth and the long prediction horizon of three jointly make the prediction task more challenging, and TPA-LSTM's accuracy improvement is not as big as in the less challenging 4G cases and shorter prediction horizons.  

\begin{table}

\caption{Bandwidth Prediction Accuracy on 5G Driving Trace
  ~\label{tbl:Re5G}}
 \begin{center} 
\begin{tabular}{lllll}\hline
5G\_Driving &   Mean: & 27.5797    & STD:  & 51.2763 \\\hline
Horizon = 1 & RLS        &  RF & LSTM       & TPA-LSTM   \\\hline
RMSE        & 30.8272 & 25.2600    & 25.1736 & {\bf24.8120} \\
MAE         & 12.6288 & 10.7901    & 10.0174 & {\bf9.0615}  \\
CORR        & 0.8016     & 0.8713        & 0.8713     & {\bf0.8771}     \\\hline
            &            &               &            &            \\\hline
Horizon = 2 & RLS        &  RF & LSTM       & TPA-LSTM   \\\hline
RMSE        & 38.3078 & 34.5933    & 34.4389 & {\bf33.6579} \\
MAE         & 16.1722 & 16.7128    & 15.7254 & {\bf13.5419} \\
CORR        & 0.6734     & 0.7503        & 0.7476     & {\bf0.7605}     \\\hline
            &            &               &            &            \\\hline
Horizon = 3 & RLS        &  RF & LSTM       & TPA-LSTM   \\\hline
RMSE        & 42.9236 & {\bf39.2689}    & 39.9783 & 40.2299 \\
MAE         & 18.4830 & 20.4666    & 19.9049 & {\bf17.0739} \\
CORR        & 0.5680     & {\bf0.6547}        & 0.6506     & 0.6331    \\\hline
\end{tabular}
\end{center}
\end{table}

\subsection{LTE/5G Handoff Prediction}
\begin{figure}

     {%
       \includegraphics[width=3.3in]{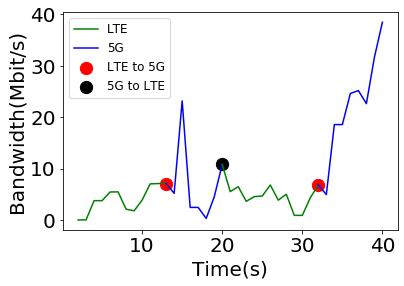}
     }
     \hfill
     \caption{Sample Trace with 5G/LTE Handoffs}
     \label{fig:5g handoff}
\end{figure}

It is expected that 5G and 4G/LTE will coexist for a long time in the transition phase. A mobile device will frequently switch between 4G and 5G access. Figure~\ref{fig:5g handoff} shows a sample bandwidth trace of $40$ seconds in the UCC 5G driving dataset. The handoffs between 4G and 5G are marked in red circles. The handoffs occur frequently, and the device stays within one mode for only around 10 seconds before switching to the other mode, largely due to the driving speed of $43.65$ km/h. Table~\ref{tbl:var 5g} shows the handoff statistics of the whole driving trace. The average sojourn time with each mode after a handoff is less than $100$ seconds and the frequency of handoff is as high as $265$  times in a trace of $18,047$ seconds.

\begin{table*}
\centering  
\caption{4G/5G Handoff Statistics in UCC 5G Driving Trace~\label{tbl:var 5g}} 
\begin{tabular}{lcccccc}  
\hline
  &$5G \rightarrow LTE$	&$LTE \rightarrow 5G$ & Avg. 5G Sojourn & Avg. LTE Sojourn &  STD 5G Sojourn &  STD LTE Sojourn\\ \hline  
  &132(times)	&133(times)	 &81.48(s)  &54.21(s) & 147.25(s) & 104.74(s)\\\hline
\end{tabular}
\end{table*}

As witnessed in Table~\ref{tbl:stat 5g}, there are apparent bandwidth differences between 5G and 4G accesses. In addition to high bandwidth, 5G is also designed with other new QoS targets, such as ultra-reliable and low-latency communication (URLLC)~\cite{ji2018ultra}, which is completely absent from LTE networks. 5G is expected to have a QoS leap. In many application scenarios, high data throughput is not the only QoS consideration. For example, in remote surgery and autonomous driving, the key is that the command and feedback signals should be sent and received with low latency. In autonomous driving, if the control signals cannot be sent and received in time, it will lead to critical consequences. If one can estimate whether a mobile device will have 5G access in the near future, it will provide valuable information for many applications to adapt their operations. Going back to the autonomous driving example, if one can predict the availability of 5G access, the autonomous driving application can plan ahead for ``normal strategy" or ``conservative strategy"  that works with short or long latencies. For a mobile AR application, the low latency of 5G access can support more frequent user interaction with virtual objects and more real-time feedback, meanwhile the high throughput of 5G access can facilitate data-intensive computation offload to edge servers. We now study 4G/5G handoff prediction.

\subsubsection{Handoff Prediction Problem}
\noindent
Let $m(t)$ be an indicator variable representing whether the current access mode is 5G or not. A handoff from 5G to 4G/LTE occurs at $t$ if $m(t-1)=1$ and $m(t)=0$. Similarly,  a handoff from 4G/LTE to 5G occurs at $t$ if $m(t-1)=0$ and $m(t)=1$. To predict handoff at a future time $t+\tau$, we simply need to estimate:
\begin{equation}
\hat m (t+\tau)=\mathbf G_d \left(\{{X(k)}, k=t-w+1, \cdots, t\}\right), 
\end{equation}
where $w$ is the history window size, $X(k)$ is the past measurements, including $m(k)$. This can be studied as a binary classification problem.

Due to the frequent handoffs back-forth between 5G and 4G, $m(t)$ oscillates between $0$ and $1$ during the transient period. We introduce a continuous version of the handoff problem. Namely, we introduce a continuous variable $\rho(t)$ as the fraction of time that 5G access is used in a future window $[t, t+\Delta-1]$, i.e., $\rho(t)=\sum_{k=t}^{t+\Delta-1} m(k)/\Delta$. We can then estimate the probability that the device will have 5G access in a future time window starting at $t+\tau$ as:
\begin{equation}
\hat \rho (t+\tau)=\mathbf G_c \left(\{{X(k)}, k=t-w-1, \cdots, t\}\right).  
\end{equation} 
This can be studied as a regression problem. 

\subsubsection{GBM-based Binary Handoff Prediction}

\noindent Gradient Boosting Machine (GBM)~\cite{friedman2001greedy} is an Ensemble Learning method for prediction problems. The main idea of GBM is to ensemble many weak prediction models, generating many sequential decision trees, to build a strong prediction model. To be specific, for inputs that consist of many parameters, every GBM decision tree would generate an output. Outputs of all decision trees are fused to generate the final output. Compared with the other traditional machine learning classification models, Ensemble Learning algorithms are more robust against over-fitting when the number of input features is large. Typically, there are two categories, {\it Gradient Boosting Classifier (GBC)}, which is to solve classification problems, and {\it Gradient Boosting Regressor (GBR)}, which is to solve regression problems. In our LTE/5G handoff Prediction, we apply GBC to predict binary handoff events (i.e. whether network mode will change or not), and apply GBR to predict the continuous version of handoffs, (i.e., the chance of 5G access in the near future). We use the Scikit-Learn~\cite{pedregosa2011scikit} library to build our GBC and GBR models.

\textbf{Data Pre-processing:}
\noindent For handoff prediction, we use the same UCC 5G driving dataset~\cite{raca2020beyond}. We use the past data from the last 5 seconds, i.e., $w$ = 5, to predict whether the access mode will change in the near future. Since handoff prediction is for applications to adjust their operations, to give applications additional time to prepare for upcoming changes, we mark a handoff if the access mode will switch within the next three seconds. Based on that, we create the dataset to train and test our handoff prediction models: extracting all $750$ input-output pairs where handoffs happened, and randomly picking $750$ input-output pairs where there is no handoff. {Among the 750 negative samples without handoff, half of them are temporally close to the positive handoff samples, and the remaining half are far from the handoff samples.}

\textbf{Features for Handoff Prediction:} 

\noindent As discussed, we use data from the last 5 seconds for handoff prediction. For each second, we first consider all the features from our bandwidth prediction experiments. Here, instead of the calculated \texttt{Cell-handoff} feature, we use the raw \texttt{CellID}. Additionally, we further process the raw \texttt{NetworkMode} feature in the previous steps to inform the model whether the access mode has switched in the recent past. Finally, we concatenate the second-wise features from 5 seconds together. This results in 13 x 5 = 65 features in total. Other than using the entire features, we also try two other feature combination sets to predict handoffs: 1) only use the bandwidth features \texttt{UL\_bitrate} and \texttt{DL\_bitrate} as input features; 2) use all features except for \texttt{UL\_bitrate} and \texttt{DL\_bitrate} bandwidth features. Table~\ref{tbl:diff_fea_handoff} summarized the different feature sets for our handoff prediction experiments. We also tried to add statistical features, such as average, variance, median, etc., to the raw features, but did not achieve significant improvement. We just use raw features in the rest of the study.

\begin{table}
\centering  
\caption{Different feature sets for our handoff prediction experiments. The BW feature set contains bandwidth features only, and the w.o. BW set contains all features except for the bandwidth features. 
~\label{tbl:diff_fea_handoff}}
 \begin{tabular}{lcc}  %
\hline
Features & \# features per second & total features\\ \hline  %
{\bf All} & 13  & 65\\   \hline 
{\bf BW Only} & 2 & 10\\   \hline 
{\bf w.o. BW}  & 11 & 55 \\  \hline
\end{tabular}
\end{table}

\textbf{Parameter Tuning:} In Gradient Boosting Classifier, we mainly tune four parameters:  \texttt{n\_estimators},  \texttt{learning\_rate},  \texttt{max\_features}, and  \texttt{max\_depth} in the training stage. 

For example, for \texttt{learning rate}, we tried different values. For each candidate learning rate, we do 5-fold validation, where we will get 5 disjoint train-validation set split. For each split, we compute the prediction accuracy when the model built on the training set is run on the testset. We then compute the average accuracy of these five disjoint configurations as the performance of this specific learning rate.  As we can see at Table~\ref{tbl:Accuracy_allwithBand}, for experiments that use all features, when the learning rate equals 0.04, we can get the highest average accuracy, which is 0.767. So, we set the learning rate as $0.04$. Similarly, the best learning rate for only using bandwidth features is obtained as $0.1$, and the best learning rate for the feature set without bandwidth features is $0.0475$. Similarly, we set the other parameters as: \texttt{n\_estimators} = 500, \texttt{max\_features} = 65 ($10$ for BW Only feature set, and $55$ for w.o BW feature set);  and \texttt{max\_depth} = 8.

\begin{figure}
\centering
\includegraphics[height=2.5in]{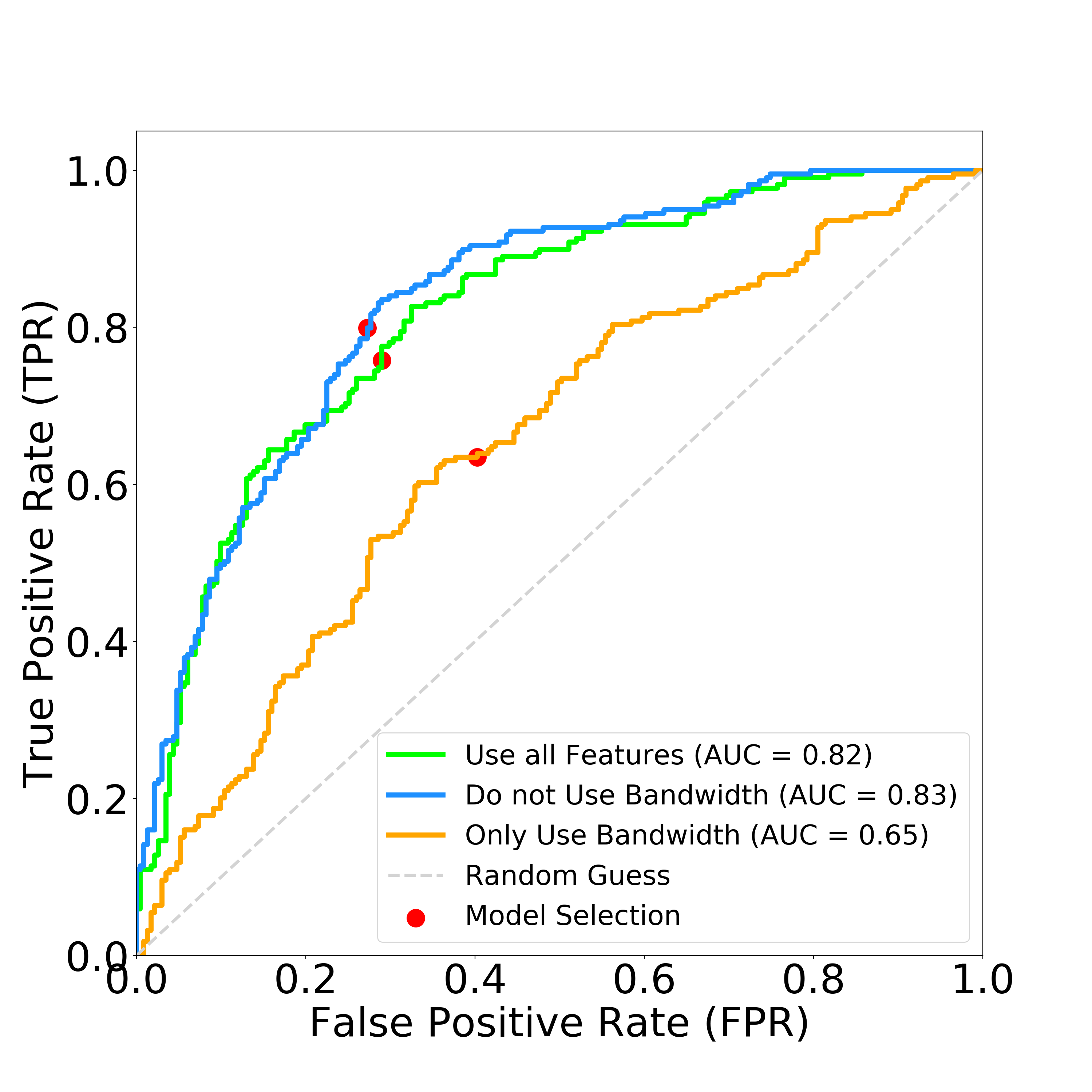}
\caption{ROC Curve for Unified Model~\label{fig:Multiple_VS}}
\end{figure}

\begin{figure}
\centering
\includegraphics[height=2.5in]{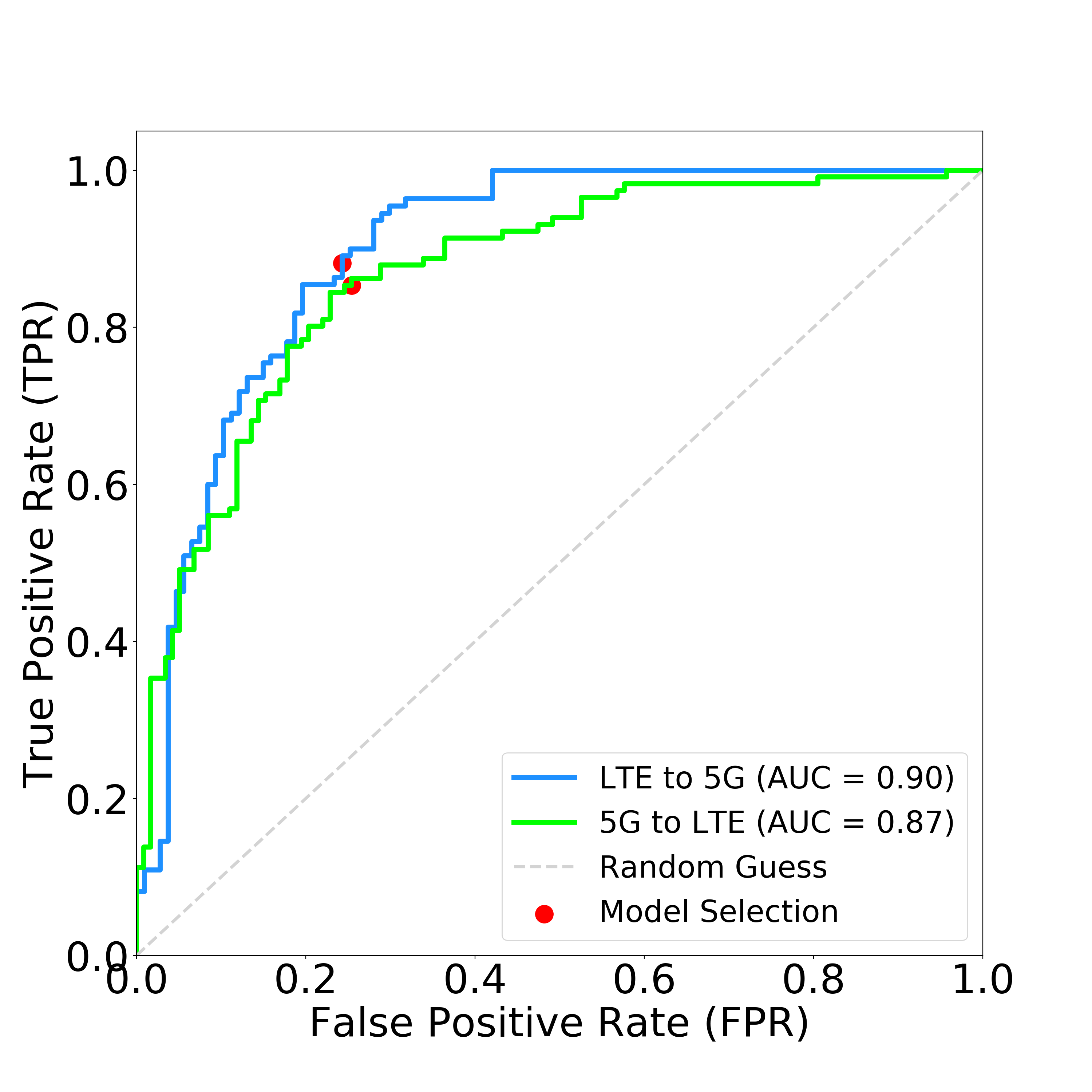}
\caption{ROC Curve for Separated Models~\label{fig:Multiple_VS1}}
\end{figure}

\begin{figure}
\centering

\includegraphics[height=2.5in]{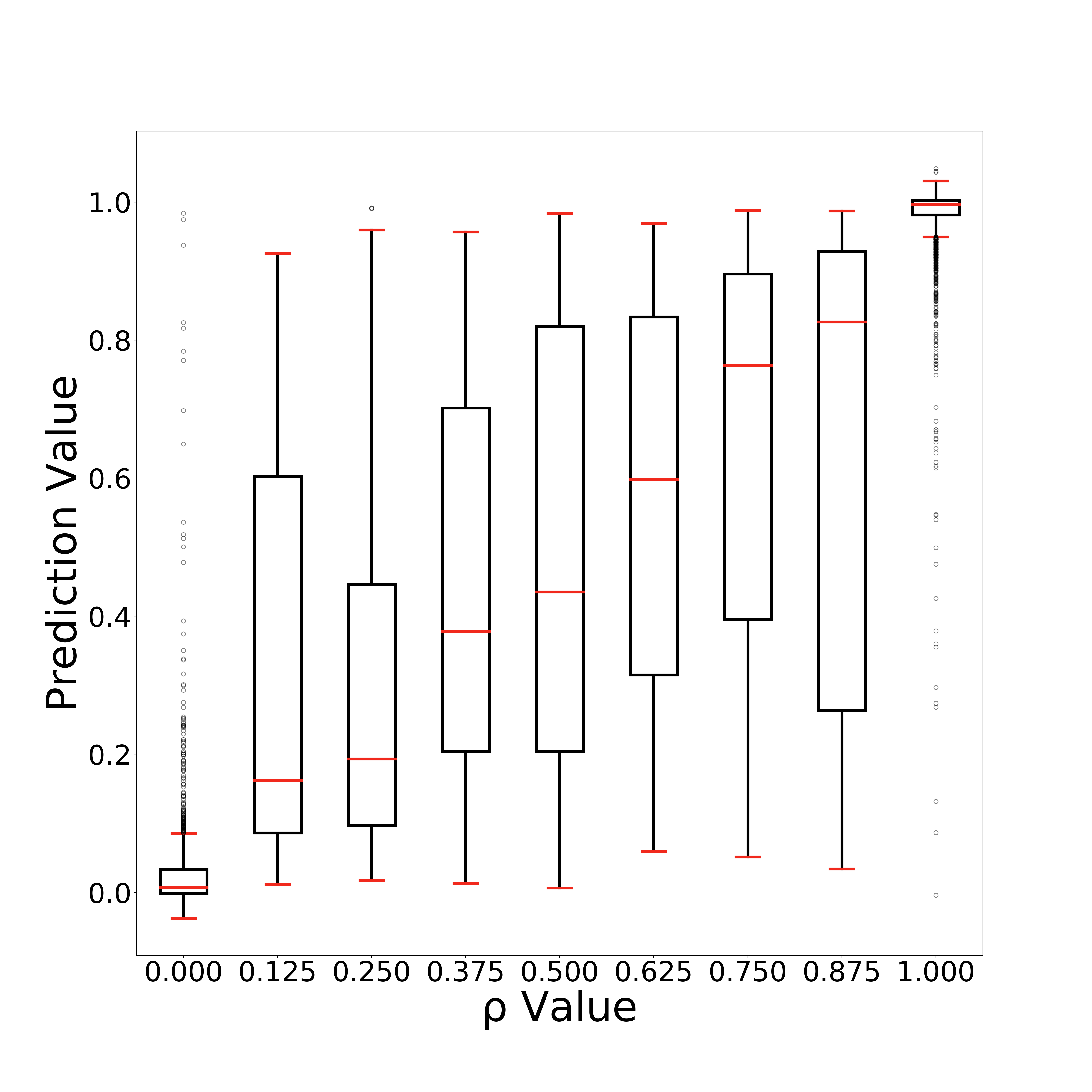}
\caption{Box Plots for Prediction Accuracy at Different $\rho$ Values~\label{fig:Box_p}}
\end{figure}

\begin{table}
\centering  %
\caption{Accuracy on Validation Set with Different Learning Rates (All features)  
  ~\label{tbl:Accuracy_allwithBand}}
\begin{tabular}{lccccc}  %
\hline
Learning Rate& 0.01& {\bf 0.04} & 0.0475 & 0.05 & 0.0525\\   %
Average Accuracy & {0.737} & {\bf 0.767} & 0.751 & 0.766 & 0.763\\   \hline 
Learning Rate & 0.055 & 0.1 & 0.25 & 0.5 & 0.75\\   %
Average Accuracy & 0.756 & 0.759 & 0.753 & 0.740 & 0.724\\   \hline 
\end{tabular}
\end{table}

\textbf{Results:}
\noindent For the dataset, 70\% of the data is in the training set, the rest 30\% is in the test set.  The confusion matrices for three feature sets are compared in Table~\ref{tbl:conf_ALL}.  ``0" represents no handoff, and ``1" represents handoff. 
Table~\ref{tbl:accuray_ALL} reports the True Positive Rate (TPR), False Positive Rate (FPR), Accuracy, F1 Score for each feature set.

\begin{table}
\centering  %
\caption{Confusion Matrices,  the values within each cell are for different feature combinations: All/BW/w.o. BW
  ~\label{tbl:conf_ALL}}
 \begin{tabular}{lcc}  %
\hline
 & Predicted 0& Predicted 1\\ \hline  %
Actual 0 & 164/138/{\bf 168} (TP)  & 67/93/{\bf 63} (FP)\\   \hline 
Actual 1 & 53/80/{\bf 44} (FN) & 166/139/{\bf 175} (TP)\\   \hline 
\end{tabular}
\end{table}

\begin{table*}
\centering 
\caption{Prediction Performance of Different Feature Combinations
  ~\label{tbl:accuray_ALL}}

\begin{tabular}{clccccc}  %
\hline
Features & TPR & FPR & Precision & Recall& Accuracy & F1\\ \hline  %
{\bf All} & 0.758 & 0.290 & 0.712 & 0.758 & 0.733 & 0.735\\   \hline 
{\bf BW Only} & 0.635 & 0.403 & 0.599 & 0.635 & 0.616 & 0.616\\   \hline 
{\bf w.o. BW} & 0.799 & 0.273 & 0.735 & 0.799 & 0.762 & 0.766\\   \hline 
\end{tabular}
\end{table*}


\noindent We also draw Receiver Operating Characteristics (ROC) curves and compute the Area Under the Curve (AUC) for GBC with three feature sets in Figure~\ref{fig:Multiple_VS}. We also mark the selected TP/FP trade-off point on the curve in red. For the bandwidth-only feature set, the AUC is only 0.65, which means this model has poor discrimination. When we use all features or all except for bandwidth features, the AUC is 0.82 and 0.83, respectively, which means these two models have excellent discrimination.
It is clear that only using bandwidth features cannot predict the handoff well. Other channel quality and context features can significantly improve the accuracy. Notice that the green curve and the blue curve are very close. This implies that adding bandwidth information into the feature set does not really improve accuracy (made it slightly worse instead).  However, in our previous correlation analysis, bandwidth and mode switch have a high correlation with handoffs. {\it This suggests that correlation does not necessarily translate into causality. In our setting, handoffs are triggered by channel quality changes, resulted from device mobility, and bandwidth variation is also a consequence of channel quality changes and handoffs. Therefore, it is important to look into channel and context information when predicting handoffs and bandwidth variations in near future.}

\subsubsection{Separated $4G \rightarrow 5G$ and $5G \rightarrow 4G$ Prediction Models}

\noindent All the experiments above only consider whether network mode will switch, no matter the switch is from 4G to 5G or 5G to 4G. It is expected that the two types of handoffs follow very different patterns. Now we build separated models for two types of handoffs. We divide the handoff datasets into two subsets, one with all samples where the current access mode is 5G, the other one with all samples where the current access mode is 4G. Then we train separated GBC models using 5-fold cross validation and optimal learning rate tuning. The prediction results are shown in Table~\ref{tbl:Accuracy2}.  We also draw Receiver Operating Characteristics (ROC) curve and compute the Area Under the Curve (AUC) for these two models in Figure~\ref{fig:Multiple_VS1}.  As expected, while the prediction accuracies for the two types of handoffs are similar, the performance of the separated models are better than the single model for both types of handoffs. 
{\it This demonstrates that customized handoff models can better mine the latent characteristics of each type of handoff for more accurate prediction.}

\begin{table*}
\caption{Performance of Separated Handoff Prediction Models
  ~\label{tbl:Accuracy2}}
\begin{tabular}{clccccc}  %
\hline
Handoff & TPR & FPR & Precision & Recall& Accuracy & F1 \\ \hline  %
$5G$ to $4G$ & 0.853 & 0.254 & 0.767 & 0.853 & 0.799 & 0.808\\   \hline 
$4G$ to $5G$ & 0.882 & 0.243 & 0.789 & 0.882 & 0.820 & 0.833\\   \hline 
\end{tabular}
\end{table*}

\noindent

\subsubsection{GBR-based Continuous Handoff Prediction}
For the continuous version of the handoff prediction, we use all the features from the past 10 seconds as input, i.e., window size $w=10$, and set the future window size $\Delta=8$. Then $\rho(t)$ is the fraction of time that the device will have 5G access within $[t, t+8]$. For example, $\rho$ = 1 means the access mode in every second of the next 8 seconds is 5G. We draw boxplots of the predicted $\hat \rho$ values for all ground-truth $\rho$ values ranging from $0$ to $1$, in Figure~\ref{fig:Box_p}. Most test samples have $\rho$ value of either $0$ or $1$. As we can see, the prediction errors and their variances for $\rho$ = 0 or $\rho$ = 1 are small, reflected by the narrow boxes centered around the true value. The variances of predictions for samples with $\rho$ value between 0 and 1 is large, reflected by the wide boxes. In general, the prediction median is in line with the ground truth value ranging from $0$ to $1$. The RMSE between the predicted value and the ground-truth is $0.109$. {\it This suggests that GBR can predict well the probability/fraction of 5G access in the near future. Such prediction can provide valuable hints for applications to adjust their operations based on the expected QoS metrics in each access mode.}

\section{Conclusion} \label{sec:conclusion}

In this paper, we studied the realtime mobile bandwidth and handoff prediction problem using 4G and 5G traces. For fixed-route mobility scenarios, we collected long consecutive traces with rich features, and demonstrated that LSTM, TPA-LSTM in particular, can effectively mine temporal patterns in channel and context information for accurate future bandwidth prediction. For 4G \& 5G co-existing networks, we proposed a new 5G/4G handoff prediction problem to mobile networking and multimedia system community. For binary and continuous 5G/4G handoff prediction problems, we developed GBM-based classification and regression models to achieve $80+\%$ prediction accuracy. As future work, we will collect a more extensive 5G dataset in NYC MTA system to further improve our prediction models. We also plan to integrate the developed prediction models into low-latency live video streaming and AR application designs to demonstrate how realtime bandwidth and handoff predictions can improve application QoE.

\printcredits



\bibliography{cas-refs}

\begin{thebibliography}{33}
\expandafter\ifx\csname natexlab\endcsname\relax\def\natexlab#1{#1}\fi
\providecommand{\url}[1]{\texttt{#1}}
\providecommand{\href}[2]{#2}
\providecommand{\path}[1]{#1}
\providecommand{\DOIprefix}{doi:}
\providecommand{\ArXivprefix}{arXiv:}
\providecommand{\URLprefix}{URL: }
\providecommand{\Pubmedprefix}{pmid:}
\providecommand{\doi}[1]{\href{http://dx.doi.org/#1}{\path{#1}}}
\providecommand{\Pubmed}[1]{\href{pmid:#1}{\path{#1}}}
\providecommand{\bibinfo}[2]{#2}
\ifx\xfnm\relax \def\xfnm[#1]{\unskip,\space#1}\fi
\bibitem[{Al-Falahy and Alani(2019)}]{al2019millimetre}
\bibinfo{author}{Al-Falahy, N.}, \bibinfo{author}{Alani, O.Y.},
  \bibinfo{year}{2019}.
\newblock \bibinfo{title}{Millimetre wave frequency band as a candidate
  spectrum for 5g network architecture: A survey}.
\newblock \bibinfo{journal}{Physical Communication} \bibinfo{volume}{32},
  \bibinfo{pages}{120--144}.
\bibitem[{Brown et~al.(1992)Brown, Hwang et~al.}]{brown1992introduction}
\bibinfo{author}{Brown, R.G.}, \bibinfo{author}{Hwang, P.Y.}, et~al.,
  \bibinfo{year}{1992}.
\newblock \bibinfo{title}{Introduction to random signals and applied Kalman
  filtering}. volume~\bibinfo{volume}{3}.
\newblock \bibinfo{publisher}{Wiley New York}.
\bibitem[{Burbank(2019)}]{5g4glatency}
\bibinfo{author}{Burbank, J.L.}, \bibinfo{year}{2019}.
\newblock \bibinfo{title}{5g vs 4g latency}.
\newblock \URLprefix
  \url{https://futurenetworks.ieee.org/images/files/pdf/FirstResponder/2019/Jack-Burbank.pdf}.
\bibitem[{Friedman(2001)}]{friedman2001greedy}
\bibinfo{author}{Friedman, J.H.}, \bibinfo{year}{2001}.
\newblock \bibinfo{title}{Greedy function approximation: a gradient boosting
  machine}.
\newblock \bibinfo{journal}{Annals of statistics} ,
  \bibinfo{pages}{1189--1232}.
\bibitem[{Ge et~al.(2009)Ge, Wen, Zheng, Lu and Wang}]{ge2009history}
\bibinfo{author}{Ge, H.}, \bibinfo{author}{Wen, X.}, \bibinfo{author}{Zheng,
  W.}, \bibinfo{author}{Lu, Z.}, \bibinfo{author}{Wang, B.},
  \bibinfo{year}{2009}.
\newblock \bibinfo{title}{A history-based handover prediction for lte systems},
  in: \bibinfo{booktitle}{2009 International Symposium on Computer Network and
  Multimedia Technology}, \bibinfo{organization}{IEEE}. pp.
  \bibinfo{pages}{1--4}.
\bibitem[{Gers et~al.(1999)Gers, Schmidhuber and Cummins}]{gers1999learning}
\bibinfo{author}{Gers, F.A.}, \bibinfo{author}{Schmidhuber, J.},
  \bibinfo{author}{Cummins, F.}, \bibinfo{year}{1999}.
\newblock \bibinfo{title}{Learning to forget: Continual prediction with lstm} .
\bibitem[{Haykin(2008)}]{haykin2008adaptive}
\bibinfo{author}{Haykin, S.}, \bibinfo{year}{2008}.
\newblock \bibinfo{title}{Adaptive filter theory. pearson education india}, in:
  \bibinfo{booktitle}{27th Annual International Conference of the Engineering
  in Medicine and Biology Society}, \bibinfo{organization}{IEEE Press}. pp.
  \bibinfo{pages}{1212--1215}.
\bibitem[{He et~al.(2007)He, Dovrolis and Ammar}]{he2007predictability}
\bibinfo{author}{He, Q.}, \bibinfo{author}{Dovrolis, C.},
  \bibinfo{author}{Ammar, M.}, \bibinfo{year}{2007}.
\newblock \bibinfo{title}{On the predictability of large transfer tcp
  throughput}.
\newblock \bibinfo{journal}{Computer Networks} \bibinfo{volume}{51},
  \bibinfo{pages}{3959--3977}.
\bibitem[{Hochreiter and Schmidhuber(1997)}]{hochreiter1997long}
\bibinfo{author}{Hochreiter, S.}, \bibinfo{author}{Schmidhuber, J.},
  \bibinfo{year}{1997}.
\newblock \bibinfo{title}{Long short-term memory}.
\newblock \bibinfo{journal}{Neural computation} \bibinfo{volume}{9},
  \bibinfo{pages}{1735--1780}.
\bibitem[{Ji et~al.(2018)Ji, Park, Yeo, Kim, Lee and Shim}]{ji2018ultra}
\bibinfo{author}{Ji, H.}, \bibinfo{author}{Park, S.}, \bibinfo{author}{Yeo,
  J.}, \bibinfo{author}{Kim, Y.}, \bibinfo{author}{Lee, J.},
  \bibinfo{author}{Shim, B.}, \bibinfo{year}{2018}.
\newblock \bibinfo{title}{Ultra-reliable and low-latency communications in 5g
  downlink: Physical layer aspects}.
\newblock \bibinfo{journal}{IEEE Wireless Communications} \bibinfo{volume}{25},
  \bibinfo{pages}{124--130}.
\bibitem[{Jiang et~al.(2014)Jiang, Sekar and Zhang}]{jiang2014improving}
\bibinfo{author}{Jiang, J.}, \bibinfo{author}{Sekar, V.},
  \bibinfo{author}{Zhang, H.}, \bibinfo{year}{2014}.
\newblock \bibinfo{title}{Improving fairness, efficiency, and stability in
  http-based adaptive video streaming with festive}.
\newblock \bibinfo{journal}{IEEE/ACM Transactions on Networking (ToN)}
  \bibinfo{volume}{22}, \bibinfo{pages}{326--340}.
\bibitem[{Kim et~al.(2007)Kim, Yang, Lee, Park and Shin}]{kim2007mobility}
\bibinfo{author}{Kim, T.H.}, \bibinfo{author}{Yang, Q.}, \bibinfo{author}{Lee,
  J.H.}, \bibinfo{author}{Park, S.G.}, \bibinfo{author}{Shin, Y.S.},
  \bibinfo{year}{2007}.
\newblock \bibinfo{title}{A mobility management technique with simple handover
  prediction for 3g lte systems}, in: \bibinfo{booktitle}{2007 IEEE 66th
  vehicular technology conference}, \bibinfo{organization}{IEEE}. pp.
  \bibinfo{pages}{259--263}.
\bibitem[{Kingma and Ba(2014)}]{kingma2014adam}
\bibinfo{author}{Kingma, D.P.}, \bibinfo{author}{Ba, J.}, \bibinfo{year}{2014}.
\newblock \bibinfo{title}{Adam: A method for stochastic optimization}.
\newblock \bibinfo{journal}{arXiv preprint arXiv:1412.6980} .
\bibitem[{Kurdoglu et~al.(2016)Kurdoglu, Liu, Wang, Shi, Gu and
  Lyu}]{kurdoglu2016real}
\bibinfo{author}{Kurdoglu, E.}, \bibinfo{author}{Liu, Y.},
  \bibinfo{author}{Wang, Y.}, \bibinfo{author}{Shi, Y.}, \bibinfo{author}{Gu,
  C.}, \bibinfo{author}{Lyu, J.}, \bibinfo{year}{2016}.
\newblock \bibinfo{title}{Real-time bandwidth prediction and rate adaptation
  for video calls over cellular networks}, in: \bibinfo{booktitle}{Proceedings
  of the 7th International Conference on Multimedia Systems},
  \bibinfo{organization}{ACM}. p.~\bibinfo{pages}{12}.
\bibitem[{Lee et~al.(2020)Lee, Lee, Lee, Sathyanarayana, Lim, Lee, Zhu,
  Ramakrishnan, Grunwald, Lee et~al.}]{lee2020perceive}
\bibinfo{author}{Lee, J.}, \bibinfo{author}{Lee, S.}, \bibinfo{author}{Lee,
  J.}, \bibinfo{author}{Sathyanarayana, S.D.}, \bibinfo{author}{Lim, H.},
  \bibinfo{author}{Lee, J.}, \bibinfo{author}{Zhu, X.},
  \bibinfo{author}{Ramakrishnan, S.}, \bibinfo{author}{Grunwald, D.},
  \bibinfo{author}{Lee, K.}, et~al., \bibinfo{year}{2020}.
\newblock \bibinfo{title}{Perceive: deep learning-based cellular uplink
  prediction using real-time scheduling patterns}, in:
  \bibinfo{booktitle}{Proceedings of the 18th International Conference on
  Mobile Systems, Applications, and Services}, pp. \bibinfo{pages}{377--390}.
\bibitem[{Liaw et~al.(2002)Liaw, Wiener et~al.}]{liaw2002classification}
\bibinfo{author}{Liaw, A.}, \bibinfo{author}{Wiener, M.}, et~al.,
  \bibinfo{year}{2002}.
\newblock \bibinfo{title}{Classification and regression by randomforest}.
\newblock \bibinfo{journal}{R news} \bibinfo{volume}{2},
  \bibinfo{pages}{18--22}.
\bibitem[{Mei et~al.(2019)Mei, Hu, Cao, Liu, Han, Li and Li}]{mei2019realtime}
\bibinfo{author}{Mei, L.}, \bibinfo{author}{Hu, R.}, \bibinfo{author}{Cao, H.},
  \bibinfo{author}{Liu, Y.}, \bibinfo{author}{Han, Z.}, \bibinfo{author}{Li,
  F.}, \bibinfo{author}{Li, J.}, \bibinfo{year}{2019}.
\newblock \bibinfo{title}{Realtime mobile bandwidth prediction using lstm
  neural network}, in: \bibinfo{booktitle}{International Conference on Passive
  and Active Network Measurement}, \bibinfo{organization}{Springer}. pp.
  \bibinfo{pages}{34--47}.
\bibitem[{Mei et~al.(2020)Mei, Hu, Cao, Liu, Han, Li and Li}]{mei2020realtime}
\bibinfo{author}{Mei, L.}, \bibinfo{author}{Hu, R.}, \bibinfo{author}{Cao, H.},
  \bibinfo{author}{Liu, Y.}, \bibinfo{author}{Han, Z.}, \bibinfo{author}{Li,
  F.}, \bibinfo{author}{Li, J.}, \bibinfo{year}{2020}.
\newblock \bibinfo{title}{Realtime mobile bandwidth prediction using lstm
  neural network and bayesian fusion}.
\newblock \bibinfo{journal}{Computer Networks} \bibinfo{volume}{182},
  \bibinfo{pages}{107515}.
\bibitem[{Mirza et~al.(2007)Mirza, Sommers, Barford and Zhu}]{mirza2007machine}
\bibinfo{author}{Mirza, M.}, \bibinfo{author}{Sommers, J.},
  \bibinfo{author}{Barford, P.}, \bibinfo{author}{Zhu, X.},
  \bibinfo{year}{2007}.
\newblock \bibinfo{title}{A machine learning approach to tcp throughput
  prediction}, in: \bibinfo{booktitle}{ACM SIGMETRICS Performance Evaluation
  Review}, \bibinfo{organization}{ACM}. pp. \bibinfo{pages}{97--108}.
\bibitem[{Narayanan et~al.(2020)Narayanan, Ramadan, Carpenter, Liu, Liu, Qian
  and Zhang}]{narayanan2020first}
\bibinfo{author}{Narayanan, A.}, \bibinfo{author}{Ramadan, E.},
  \bibinfo{author}{Carpenter, J.}, \bibinfo{author}{Liu, Q.},
  \bibinfo{author}{Liu, Y.}, \bibinfo{author}{Qian, F.},
  \bibinfo{author}{Zhang, Z.L.}, \bibinfo{year}{2020}.
\newblock \bibinfo{title}{A first look at commercial 5g performance on
  smartphones}, in: \bibinfo{booktitle}{Proceedings of The Web Conference
  2020}, pp. \bibinfo{pages}{894--905}.
\bibitem[{Pedregosa et~al.(2011)Pedregosa, Varoquaux, Gramfort, Michel,
  Thirion, Grisel, Blondel, Prettenhofer, Weiss, Dubourg
  et~al.}]{pedregosa2011scikit}
\bibinfo{author}{Pedregosa, F.}, \bibinfo{author}{Varoquaux, G.},
  \bibinfo{author}{Gramfort, A.}, \bibinfo{author}{Michel, V.},
  \bibinfo{author}{Thirion, B.}, \bibinfo{author}{Grisel, O.},
  \bibinfo{author}{Blondel, M.}, \bibinfo{author}{Prettenhofer, P.},
  \bibinfo{author}{Weiss, R.}, \bibinfo{author}{Dubourg, V.}, et~al.,
  \bibinfo{year}{2011}.
\newblock \bibinfo{title}{Scikit-learn: Machine learning in python}.
\newblock \bibinfo{journal}{the Journal of machine Learning research}
  \bibinfo{volume}{12}, \bibinfo{pages}{2825--2830}.
\bibitem[{Raca et~al.(2020)Raca, Leahy, Sreenan and Quinlan}]{raca2020beyond}
\bibinfo{author}{Raca, D.}, \bibinfo{author}{Leahy, D.},
  \bibinfo{author}{Sreenan, C.J.}, \bibinfo{author}{Quinlan, J.J.},
  \bibinfo{year}{2020}.
\newblock \bibinfo{title}{Beyond throughput, the next generation: a 5g dataset
  with channel and context metrics}, in: \bibinfo{booktitle}{Proceedings of the
  11th ACM Multimedia Systems Conference}, pp. \bibinfo{pages}{303--308}.
\bibitem[{Raiciu et~al.(2012)Raiciu, Paasch, Barre, Ford, Honda, Duchene,
  Bonaventure and Handley}]{mptcp}
\bibinfo{author}{Raiciu, C.}, \bibinfo{author}{Paasch, C.},
  \bibinfo{author}{Barre, S.}, \bibinfo{author}{Ford, A.},
  \bibinfo{author}{Honda, M.}, \bibinfo{author}{Duchene, F.},
  \bibinfo{author}{Bonaventure, O.}, \bibinfo{author}{Handley, M.},
  \bibinfo{year}{2012}.
\newblock \bibinfo{title}{How hard can it be? designing and implementing a
  deployable multipath tcp}, in: \bibinfo{booktitle}{NSDI}.
\bibitem[{Ruan et~al.(2019)Ruan, Dias and Wong}]{ruan2019machine}
\bibinfo{author}{Ruan, L.}, \bibinfo{author}{Dias, M.P.I.},
  \bibinfo{author}{Wong, E.}, \bibinfo{year}{2019}.
\newblock \bibinfo{title}{Machine learning-based bandwidth prediction for
  low-latency h2m applications}.
\newblock \bibinfo{journal}{IEEE Internet of Things Journal}
  \bibinfo{volume}{6}, \bibinfo{pages}{3743--3752}.
\bibitem[{Shih et~al.(2019)Shih, Sun and Lee}]{shih2019temporal}
\bibinfo{author}{Shih, S.Y.}, \bibinfo{author}{Sun, F.K.},
  \bibinfo{author}{Lee, H.y.}, \bibinfo{year}{2019}.
\newblock \bibinfo{title}{Temporal pattern attention for multivariate time
  series forecasting}.
\newblock \bibinfo{journal}{Machine Learning} \bibinfo{volume}{108},
  \bibinfo{pages}{1421--1441}.
\bibitem[{Smola and Sch{\"o}lkopf(2004)}]{smola2004tutorial}
\bibinfo{author}{Smola, A.J.}, \bibinfo{author}{Sch{\"o}lkopf, B.},
  \bibinfo{year}{2004}.
\newblock \bibinfo{title}{A tutorial on support vector regression}.
\newblock \bibinfo{journal}{Statistics and computing} \bibinfo{volume}{14},
  \bibinfo{pages}{199--222}.
\bibitem[{Sun et~al.(2016)Sun, Yin, Jiang, Sekar, Lin, Wang, Liu and
  Sinopoli}]{sun2016cs2p}
\bibinfo{author}{Sun, Y.}, \bibinfo{author}{Yin, X.}, \bibinfo{author}{Jiang,
  J.}, \bibinfo{author}{Sekar, V.}, \bibinfo{author}{Lin, F.},
  \bibinfo{author}{Wang, N.}, \bibinfo{author}{Liu, T.},
  \bibinfo{author}{Sinopoli, B.}, \bibinfo{year}{2016}.
\newblock \bibinfo{title}{Cs2p: Improving video bitrate selection and
  adaptation with data-driven throughput prediction}, in:
  \bibinfo{booktitle}{Proceedings of the 2016 ACM SIGCOMM Conference},
  \bibinfo{organization}{ACM}. pp. \bibinfo{pages}{272--285}.
\bibitem[{Tian and Liu(2012)}]{tian2012towards}
\bibinfo{author}{Tian, G.}, \bibinfo{author}{Liu, Y.}, \bibinfo{year}{2012}.
\newblock \bibinfo{title}{Towards agile and smooth video adaptation in dynamic
  http streaming}, in: \bibinfo{booktitle}{Proceedings of the 8th international
  conference on Emerging networking experiments and technologies},
  \bibinfo{organization}{ACM}. pp. \bibinfo{pages}{109--120}.
\bibitem[{Wang et~al.(2016)Wang, Huang, Zhu and Zhao}]{wang2016attention}
\bibinfo{author}{Wang, Y.}, \bibinfo{author}{Huang, M.}, \bibinfo{author}{Zhu,
  X.}, \bibinfo{author}{Zhao, L.}, \bibinfo{year}{2016}.
\newblock \bibinfo{title}{Attention-based lstm for aspect-level sentiment
  classification}, in: \bibinfo{booktitle}{Proceedings of the 2016 conference
  on empirical methods in natural language processing}, pp.
  \bibinfo{pages}{606--615}.
\bibitem[{Wei et~al.(2018)Wei, Kawakami, Kanai, Katto and Wang}]{wei2018trust}
\bibinfo{author}{Wei, B.}, \bibinfo{author}{Kawakami, W.},
  \bibinfo{author}{Kanai, K.}, \bibinfo{author}{Katto, J.},
  \bibinfo{author}{Wang, S.}, \bibinfo{year}{2018}.
\newblock \bibinfo{title}{Trust: A tcp throughput prediction method in mobile
  networks}, in: \bibinfo{booktitle}{2018 IEEE Global Communications Conference
  (GLOBECOM)}, \bibinfo{organization}{IEEE}. pp. \bibinfo{pages}{1--6}.
\bibitem[{Winstein et~al.(2013)Winstein, Sivaraman and
  Balakrishnan}]{winstein2013stochastic}
\bibinfo{author}{Winstein, K.}, \bibinfo{author}{Sivaraman, A.},
  \bibinfo{author}{Balakrishnan, H.}, \bibinfo{year}{2013}.
\newblock \bibinfo{title}{Stochastic forecasts achieve high throughput and low
  delay over cellular networks}, in: \bibinfo{booktitle}{Presented as part of
  the 10th $\{$USENIX$\}$ Symposium on Networked Systems Design and
  Implementation ($\{$NSDI$\}$ 13)}, pp. \bibinfo{pages}{459--471}.
\bibitem[{Yin et~al.(2015)Yin, Jindal, Sekar and Sinopoli}]{yin2015control}
\bibinfo{author}{Yin, X.}, \bibinfo{author}{Jindal, A.},
  \bibinfo{author}{Sekar, V.}, \bibinfo{author}{Sinopoli, B.},
  \bibinfo{year}{2015}.
\newblock \bibinfo{title}{A control-theoretic approach for dynamic adaptive
  video streaming over http}, in: \bibinfo{booktitle}{ACM SIGCOMM Computer
  Communication Review}, \bibinfo{organization}{ACM}. pp.
  \bibinfo{pages}{325--338}.
\bibitem[{Yue et~al.(2018)Yue, Jin, Suh, Qin, Wang and
  Wei}]{yue2018linkforecast}
\bibinfo{author}{Yue, C.}, \bibinfo{author}{Jin, R.}, \bibinfo{author}{Suh,
  K.}, \bibinfo{author}{Qin, Y.}, \bibinfo{author}{Wang, B.},
  \bibinfo{author}{Wei, W.}, \bibinfo{year}{2018}.
\newblock \bibinfo{title}{Linkforecast: Cellular link bandwidth prediction in
  lte networks}.
\newblock \bibinfo{journal}{IEEE Transactions on Mobile Computing} ,
  \bibinfo{pages}{1582--1594}.

\end{thebibliography}

\end{document}